\documentclass[11pt]{article}

\usepackage[final]{acl}

\usepackage{times}
\usepackage{latexsym}

\usepackage[T1]{fontenc}

\usepackage[utf8]{inputenc}

\usepackage{microtype}

\usepackage{inconsolata}

\usepackage{graphicx}

\usepackage{graphicx}
\usepackage{threeparttable}
\usepackage{booktabs}
\usepackage{multirow}
\usepackage[most]{tcolorbox}
\tcbuselibrary{breakable}
\usepackage{float}
\usepackage{listings}
\usepackage{tabularx}
\usepackage{listings}
\usepackage{subcaption} 
\title{Enhancing Tool Calling in LLMs with the International Tool Calling Dataset}

\author{
  Zuoyu Zhang \\
  Shenzhen University \\
  \texttt{2310533003@email.szu.edu.cn}
  \And
  Yancheng Zhu \\
  Shenzhen University \\
  \texttt{2410673009@mails.szu.edu.cn}
}

\begin{document}
\maketitle
\begin{abstract}
Tool calling allows large language models (LLMs) to interact with external systems like APIs, enabling applications in customer support, data analysis, and dynamic content generation. While recent benchmarks have advanced tool-use research, they suffer from key limitations, including reliance on simulated or restricted APIs, limited reproducibility, and a lack of cultural and geographic diversity. To address these gaps, we introduce International Tool Calling (ITC), a large-scale, multilingual benchmark designed for realistic, globally distributed tool-calling scenarios. ITC includes 3,571 real APIs and 17,540 tool calling tasks across 20 categories and 40 countries. Experiments reveal substantial performance gaps between open- and closed-source LLMs, while fine-tuning on ITC yields significant improvements, particularly for non-English queries, enhancing cross-lingual generalization, reasoning consistency, and robustness to out-of-domain tools. ITC provides a valuable benchmark for advancing LLM robustness and performance in complex, multi-tool, and international scenarios. Dataset: \href{https://anonymous.4open.science/r/International-Tool-Calling-ITC-dataset-FAF4/}{https://anonymous.4open.science/r/International-Tool-Calling-ITC-dataset-FAF4/}.
\end{abstract}

\section{Introduction}
\label{sec:introduction}
Tool calling empowers large language models (LLMs) to interact with external systems—such as databases, APIs, and software tools—extending their capabilities beyond text generation~\citep{schick2023toolformerlanguagemodelsteach}. By invoking tools, LLMs can access real-time data, perform complex computations, and execute actions beyond their training data~\citep{nakano2021webgpt}. This functionality is essential for tasks such as automated customer support, data analysis, and dynamic content generation, where external resource integration enhances both performance and utility. As surveyed in~\citep{mialon2023augmented}, tool calling enables more sophisticated, context-aware interactions, making LLMs valuable across diverse domains.

Recent advances have led to the development of several datasets and benchmarks to improve tool-use capabilities in LLMs. Notable examples include API-BLEND~\citep{basu2024apiblendcomprehensivecorporatraining}, APIGen~\citep{Liu2024APIGenAP}, and ToolACE~\citep{liu2024toolacewinningpointsllm}, which focus on static API-based function calling across a variety of use cases. In contrast, datasets like Gorilla~\citep{patil2023gorillalargelanguagemodel} and ToolLLM~\citep{qin2023toolllmfacilitatinglargelanguage} emphasize real-world tool invocation with closed-loop execution. More complex datasets like Seal-Tools~\citep{wu2024seal}, PLUTO~\citep{huang2024planning}, SciToolBench~\citep{ma2024sciagent}, and the recent ToolHop~\citep{ye2025toolhop} explore multi-step reasoning and domain-specific tool use. Furthermore, emerging 2025 benchmarks such as ToolSandbox~\citep{lu2025toolsandbox} and CONFETTI~\citep{alkhouli2025confetti} extend evaluation to stateful, multi-turn conversational interactions, while ACEBench~\citep{chen2025acebench} offers systematic robustness assessments. Collectively, these benchmarks have significantly advanced the development of LLMs capable of interacting with external tools effectively.

However, significant challenges remain. Many existing datasets—including recent ones like Seal-Tools~\citep{wu2024seal} and ToolSandbox~\citep{lu2025toolsandbox}—rely on simulated APIs or synthetic environments, which fail to capture the complexity and variability of real-world tool usage. Others rely on real APIs but incur high costs due to paid access keys and strict usage limits. For example, ToolLLM~\citep{qin2023toolllmfacilitatinglargelanguage}, although publicly available, depends on APIs subject to quotas, key management, and usage limits, which can hinder reproducibility and limit practical deployment. In addition, some datasets are entirely inaccessible due to proprietary restrictions or other barriers. Moreover, existing benchmarks largely overlook cultural and regional diversity in tool usage. APIs are often region-specific, reflecting differences in culture, regulations, services, and user behavior across regions. The lack of geographically and culturally diverse APIs limits the generalizability of current benchmarks and underscores the need to incorporate APIs from multiple countries in tool-calling evaluations.

To overcome existing limitations in tool calling research, we present the \textbf{International Tool Calling (ITC)} dataset, specifically designed to support real-world, globally distributed tool calling scenarios. The dataset comprises 3,571 real-world APIs and 17,540 tool calling tasks—15,790 for training and 1,750 for testing—covering 20 categories across 40 countries. It includes 64.2\% global APIs—such as machine translation and international weather services—and region-specific APIs from major regions like the United States and China, along with 38 additional countries, ensuring broad geographic and functional diversity. By encompassing a wide range of single- and multi-tool tasks, ITC captures realistic challenges in tool selection, parameter specification, and cross-cultural usage, making it a comprehensive resource for evaluating and improving the performance and generalization of tool-augmented language models.

We benchmarked 16 open-source and 8 closed-source LLMs on the ITC test set, revealing substantial performance gaps across multiple metrics and highlighting common challenges in tool usage, such as handling nonexistent tools, missing parameters, and incorrect parameter generation. Fine-tuning on the full multilingual ITC dataset yields significant performance gains, particularly on non-English queries, by enhancing reasoning consistency and cross-lingual generalization, while also improving out-of-domain generation and boosting tool selection and invocation precision on external benchmarks, demonstrating ITC’s effectiveness in enhancing generalization and robustness in complex, real-world scenarios.

\section{Related Work}
\label{sec:related_work}

Existing benchmarks for enhancing LLM tool-invocation cover a variety of tasks, including API-based interactions, multi-step reasoning, and robustness evaluation. 
Datasets such as API-BLEND~\citep{basu2024apiblendcomprehensivecorporatraining}, APIGen~\citep{Liu2024APIGenAP}, and ToolACE~\citep{liu2024toolacewinningpointsllm} provide diverse APIs for training and evaluation. 
Complementing these, FuncBenchGen~\citep{maekawa2025towards} introduces a synthetic benchmark generation framework to create controllable tasks with complex dependencies. 
In terms of real-world application, Gorilla~\citep{patil2023gorillalargelanguagemodel} and ToolLLM~\citep{qin2023toolllmfacilitatinglargelanguage} improve LLM performance on API interactions. 
To address more complex scenarios, Seal-Tools~\citep{wu2024seal}, PLUTO~\citep{huang2024planning}, and SciToolBench~\citep{ma2024sciagent} have been developed, with the recent ToolHop~\citep{ye2025toolhop} specifically targeting multi-hop reasoning and chained tool execution. 
Furthermore, 2025 advancements emphasize conversational and stateful interactions; ToolSandbox~\citep{lu2025toolsandbox} and CONFETTI~\citep{alkhouli2025confetti} evaluate LLMs in multi-turn dialogues involving state dependencies, implicit goals, and goal switching. 
Regarding robustness and comprehensive evaluation, ACEBench~\citep{chen2025acebench} offers a systematic assessment across diverse scenarios, extending the efforts of RoTBench~\citep{ye2024rotbench}, StableToolBench~\citep{guo2024stabletoolbench}, ToolEyes~\citep{ye2024tooleyes}, and ToolSword~\citep{ye2024toolsword}. 
Finally, multi-modal frameworks like MLLM-Tool~\citep{wang2024tool} extend interactions to images, text, and audio.

\setlength{\tabcolsep}{4pt} 
\begin{table*}[ht!]

\vspace{-0.5em}

\centering
\small 
\begin{threeparttable}
\begin{tabular}{l c c c c c c} 
\hline
Dataset & \# Tools & Source & Tool Format & \# Tasks & Callability & \# TL \\
\hline
API-BLEND~\citep{basu2024apiblendcomprehensivecorporatraining} & - & Simulated & API Endpoints & 189,040 & $\times$ & English \\
APIGen~\citep{Liu2024APIGenAP} & 3,673 & Real & API Endpoints & 60,000 & $\surd$ & English \\
Gorilla~\citep{patil2023gorillalargelanguagemodel} & 1,645 & Real & Python Functions & 16,450 & $\surd$ & English \\
Seal-Tools~\citep{wu2024seal} & 4,076 & Simulated & API Endpoints & 14,076 & $\times$ & English \\
ToolACE~\citep{liu2024toolacewinningpointsllm} & 26,507 & Simulated & Python Functions & 11,300 & $\times$ & English \\
ToolBench~\citep{qin2023toolllmfacilitatinglargelanguage} & 16,464 & Real & API Endpoints & 126,486 & $\surd$ & English \\
RoTBench~\citep{ye2024rotbench} & 568 & ToolEyes & API Endpoints & 105 & $\surd$ & English \\
MLLM-Tool~\citep{wang2024tool} & 932 & Real & API Endpoints & 11,642 & $\surd$ & English \\
PLUTO~\citep{huang2024planning} & 2,032 & Simulated & Python Functions & 5,824 & $\times$ & English \\
SciToolBench~\citep{ma2024sciagent} & 2,446 & Simulated & Python Functions & 856 & $\times$ & English \\ 
GeoLLM-QA~\citep{singh2024evaluating} & 117 & Real & API Endpoints & 1,000 & $\times$ & English \\
INJECAGENT~\citep{zhan2024injecagent} & 17 & Simulated & Python Functions & 1,054 & $\surd$ & English \\
StableToolBench~\citep{guo2024stabletoolbench} & 16,464 & ToolBench & API Endpoints & 126,486 & $\surd$ & English \\
ToolEyes~\citep{ye2024tooleyes} & 568 & Simulated & API Endpoints & 382 & $\surd$ & English \\
ToolSword~\citep{ye2024toolsword} & 100 & Simulated & API Endpoints & 440 & $\surd$ & English \\
Hammer~\citep{lin2024hammer} & - & APIGen & Python Functions & 67,500 & $\times$ & English \\
ToolHop~\citep{ye2025toolhop} & 3912 & Simulated & Python Functions & 995 & $\surd$ & English \\
ToolSandbox~\citep{lu2025toolsandbox} & - & Simulated & Python Functions & 1,024 & $\surd$ & English \\
CONFETTI~\citep{alkhouli2025confetti} & 86 & Simulated & Python Functions & 109 & $\surd$ & English \\
ACEBench~\citep{chen2025acebench} & 4,538 & Real & API Endpoints & 2,000 & $\surd$ & English \\
\hline
\textbf{Ours} & \textbf{3,571} & \textbf{Real} & \textbf{API Endpoints} & \textbf{17,540} & \textbf{$\surd$} & \textbf{29 languages} \\
\hline
\end{tabular}
\end{threeparttable}
\caption{Summary of existing tool calling datasets. \textbf{\# Tools} denotes the number of distinct tools or APIs provided. \textbf{Source} indicates whether the tools are collected from real-world services, simulated, or derived from another benchmark such as ToolEyes. \textbf{Tool Format} distinguishes between Web/HTTP-based API endpoints and executable Python function/code calls. \textbf{\# Tasks} refers to the number of tool calling queries. \textbf{Callability} indicates whether the tools can be called in an actual runtime environment. \textbf{TL} denotes the task languages supported in task definitions (e.g., English only or multilingual).}
\label{tab:benchmarks}
\end{table*}

Table~\ref{tab:benchmarks} summarizes representative tool-calling datasets. Despite notable progress, existing benchmarks suffer from several limitations. Many rely on simulated APIs that fail to capture real-world variability; Although some datasets involving real APIs, but more than half are not publicly or freely available. Moreover, existing benchmarks largely overlook cultural and regional diversity in tool usage. To address these shortcomings, our dataset provides 3,571 real-world APIs that are publicly accessible without authentication keys, span multiple domains, and originate from 40 countries, enabling more realistic, reproducible, and globally representative evaluations of tool-calling capabilities.

\section{Dataset Curation}
\label{sec:dataset_curation}
\begin{figure}[h]
  \includegraphics[width=\columnwidth]{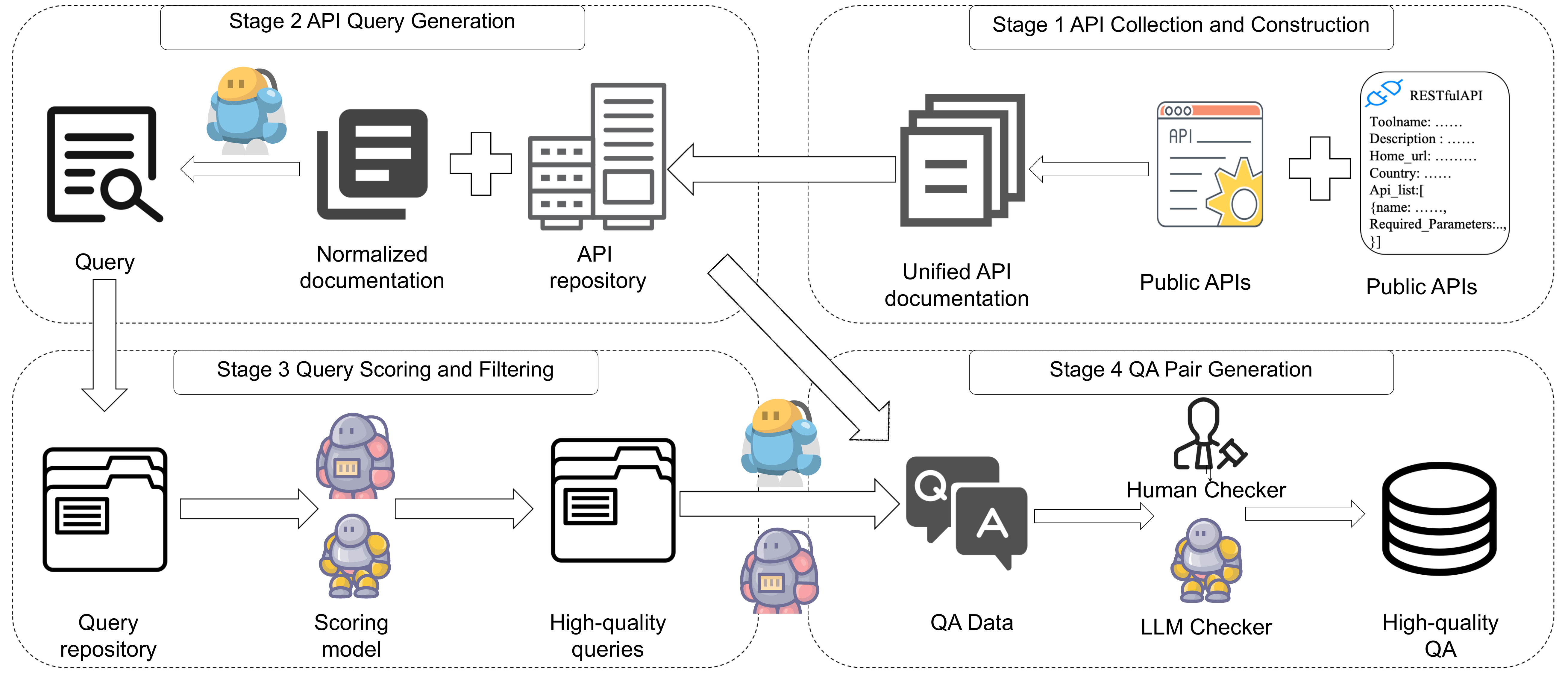}
  \caption{Dataset construction flowchart.}
  \vspace{-0.5em}
  \label{fig:process}
\end{figure}

Our pipeline is illustrated in Figure~\ref{fig:process}. F  irst collects API documentation (Stage 1), then uses GPT-4o to generate detailed API instructions (Stage 2). Next, Claude-3.5-Sonnet and Gemini-1.5-Pro refine queries for clarity and executability (Stage 3), and finally GPT-4o and Gemini-1.5-Pro generates high-quality QA pairs (Stage 4).

\subsection{API Collection and Construction}
\label{sec:API_Collection_and_Construction}



\noindent\textbf{API Collection}: We constructed a comprehensive dataset of 49,937 real-world REST APIs spanning 20 functional categories (e.g., social media, e-commerce, weather). To ensure data authenticity and traceability, all APIs were strictly collected from five specific global sources. As detailed in Table~\ref{tab:api_source_distribution}, RapidAPI~\citep{rapidapi} serves as the primary source, accounting for approximately half of the dataset (50.3\%), while the remaining portion is distributed among major regional marketplaces (e.g., Juhe Data~\citep{juhe_api}) and community-maintained repositories. This distribution ensures both the scale of commercial APIs and the diversity of open-source contributions.

\begin{table}[h]
    \centering
    
    \resizebox{\columnwidth}{!}{%
    \begin{tabular}{l l r r}
        \toprule
        \textbf{Source} & \textbf{Type} & \textbf{Count} & \textbf{Percentage} \\
        \midrule
        RapidAPI~\citep{rapidapi} & Marketplace & 25,123 & 50.3\% \\
        Juhe Data~\citep{juhe_api} & Marketplace & 11,984 & 24.0\% \\
        Public APIs~\citep{public_apis} & GitHub & 6,412 & 12.8\% \\
        Xiarou API~\citep{xiarou_api} & Marketplace & 3,541 & 7.1\% \\
        Free API~\citep{free_api} & GitHub & 2,877 & 5.8\% \\
        \midrule
        \textbf{Total} & -- & \textbf{49,937} & \textbf{100.0\%} \\
        \bottomrule
    \end{tabular}%
    }
\caption{API source distribution (total: 49,937).}
\label{tab:api_source_distribution}
\end{table}

\noindent\textbf{API Supplementation and Verification}: To ensure reliable LLM parsing, all API documents are standardized into a uniform schema covering name, description, endpoint, method, authentication, and input/output parameters (see Figure~\ref{fig:API_Format}, Appendix~\ref{sec:API_Processing}). To address incomplete specifications without relying on synthetic generation, we adopted a response-driven manual completion strategy. Specifically, for APIs with missing core metadata but active endpoints, we executed live tests using Python scripts or curl commands. We then analyzed the actual runtime responses to accurately supplement the documentation, ensuring that parameter definitions and output schemas strictly reflect real-world behaviors. Finally, correctness is rigorously verified through sample executions, and APIs with irreparable issues or connectivity failures are removed. This process produces consistent, high-quality API specifications grounded in actual execution results (example in Figure~\ref{fig:example_api}, Appendix~\ref{sec:Example_data}).

\noindent\textbf{API Filtering}: To ensure data quality and reliability, we applied a rigorous automated filtering process incorporating longitudinal stability checks. Throughout the dataset construction phase, we executed automated monitoring scripts on a weekly basis to test each API with predefined queries. This continuous evaluation enabled the removal of non-responsive APIs, those returning errors (e.g., 404 or 500), empty responses, or malformed or non-JSON outputs. This process significantly reduced the initial pool from 49,937 to 3,571 high-quality APIs (7.1\% of the original), ensuring the remaining APIs are \emph{consistently} stable and suitable for generating tool-use instructions (see examples in Figures~\ref{fig:API_Error_Response} and~\ref{fig:API_Empty_Response}, Appendix~\ref{sec:API_Processing}).

\subsection{Query Generation}
\label{sec:QueryGeneration}

We categorize tool-calling tasks into \textit{Single Tool Calling}, which invokes a single API, and \textit{Multiple Tools Calling}, coordinating several APIs. The latter includes \textit{Repeated} (same API multiple times), \textit{Parallel} (multiple APIs simultaneously), and \textit{Nested} (chained API calls) subtypes. Real-world scenarios often require multilingual and region-aware capabilities—for example, a Japanese tourist planning a trip to Lijiang in China may need local weather and travel information from a Chinese API, with both queries and responses in Japanese. To support such use cases and generate high-quality queries, we construct a manually curated seed pool consisting of 36 high-quality instances, covering all task types across diverse languages and regions.  Starting from the seed pool, queries are generated via an API-focused process. For each seed task, APIs are selected from our multilingual repository according to three principles: (1) \textbf{Geographic diversity}: include APIs from countries or regions that have fewer available APIs, such aht the dataset is not dominated by a few regions; (2) \textbf{Functional variety}: include APIs that perform similar or complementary tasks, allowing repeated, parallel, or chained calls in a scenario; 
    (3) \textbf{Disambiguation challenge}: include APIs with similar names or outputs to test whether the model can choose the correct one in context. For each API (or API set), GPT-4o generates three user queries conditioned on 1--3 task-specific seed examples (see Appendix~\ref{sec:Single_Tool_Calling_Tasks_Query_Generation} and~\ref{sec:Multi-tool_Call_Prompt}).

\subsection{Query Scoring and Filtering}
\label{sec:Query_Scoring_Filtering}

In the previous step, we obtained 44,198 generated queries, many of which suffered from unclear requirements, low relevance, non-standard language, or poor adherence to cultural context. Our query selection involved two steps: \textbf{Query Scoring} and \textbf{Query Filtering}. In the scoring step, we evaluated each query across five dimensions—Relevance, Practicality, Linguistic Applicability, Clarity, and Specificity (Appendix~\ref{sec:Scoring_dimensions})—using two independent LLMs, Claude-3.5-Sonnet and Gemini-1.5-Pro, with scores from 1 (lowest) to 5 (highest). During Query Filtering, only queries scoring above 4 from both models were retained, removing 25,830 queries (58.4\%) and leaving 18,368 candidates. To rigorously verify these remaining candidates and eliminate potential bias, we conducted a large-scale verification campaign involving 100 qualified annotators recruited via a crowdsourcing platform. These annotators were selected through strict linguistic proficiency and guideline comprehension exams. Furthermore, we implemented real-time quality control by embedding 10\% ``gold standard'' control questions, dynamically excluding workers whose accuracy dropped below 85\%. This process achieved substantial inter-annotator agreement (Fleiss' $\kappa = 0.68$, see Appendix~\ref{sec:annotation_details} for detailed statistics), with disagreements adjudicated by expert linguists. Ultimately, 17,540 queries were retained (a further 4.5\% removal), ensuring that the final dataset is highly relevant, linguistically appropriate, and reliable for downstream tasks.

\vspace{-2pt}
\subsection{Question-and-Answer Pair Generation}
\label{sec:Question-and-Answer_Pair_Generation}



To ensure the high quality of the 17,540 curated QA pairs, we adopted a task-specific generation strategy: tailored prompt templates were applied to each query based on its task classification (Single, Repeated, Parallel, or Nested Tool Calling; see Appendix~\ref{sec:QA_Generation}). To mitigate the annotation bias inherent in single-source generation, we employed a tri-model generation strategy.
Three large language models—GPT-4o, Gemini-1.5-Pro, and Claude-3.5-Sonnet—independently generated candidate answers for each query.
Each candidate was then evaluated by the other LLMs based on \emph{consistency} between reasoning and API calls, \emph{solution validity}, and \emph{linguistic quality} (see Appendix~\ref{sec:Checker}). These automated rankings were then audited by human experts, who made the final selection, particularly for high-complexity tasks. This adversarial approach effectively decoupled generation sources from evaluation, significantly reducing model-specific hallucinations and ensuring the quality of the final dataset.

\section{Data Statistics}
\label{sec:data_statistics}
Our International Tool Calling (ITC) dataset comprises 3,571 APIs and a total of 17,540 question-and-answer pairs. To create a challenging test set that evaluates generalization to unseen tools, we partitioned the data at the API level, resulting in a training set of 15,790 tasks and a test set of 1,750 tasks. This ensures the test set contains a significant portion of APIs not seen during training. In the following sections, we detail the composition of the dataset from two perspectives: APIs and Tasks.


\begin{figure}[ht]
  \includegraphics[width=\columnwidth]{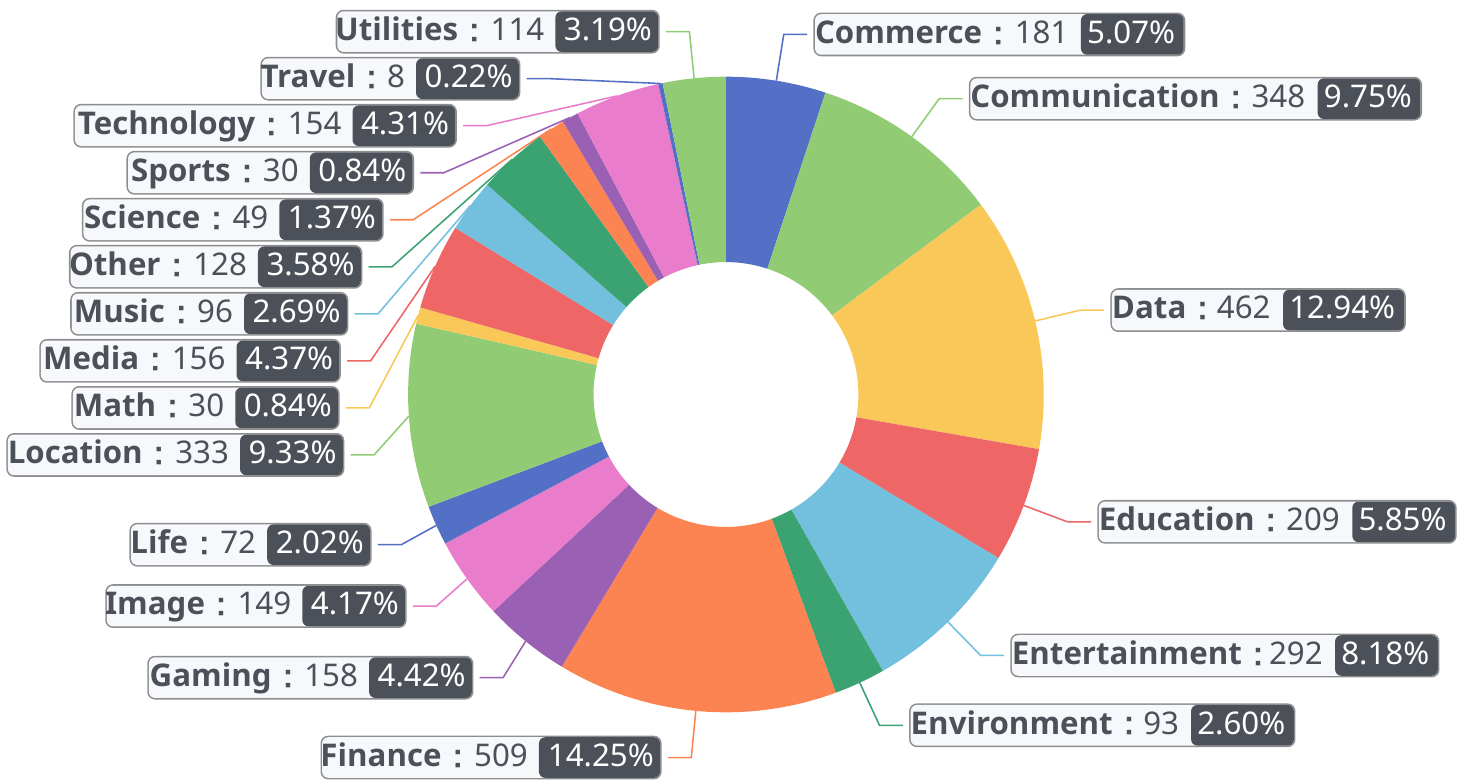}
  \caption{Category Distribution in the ITC Dataset.}
  \label{fig:tool_nums}
\end{figure}
Figure~\ref{fig:tool_nums} illustrates the distribution of APIs across 20 Category in the ITC Dataset. The largest categories are Finance (14.25\%), Data (12.9\%), Communication (9.75\%), and Entertainment (8.18\%). Conversely, the smallest categories include Travel (0.22\%), Math (0.84\%), and Sports (0.84\%).

\begin{figure}[ht]
\includegraphics[width=\linewidth]{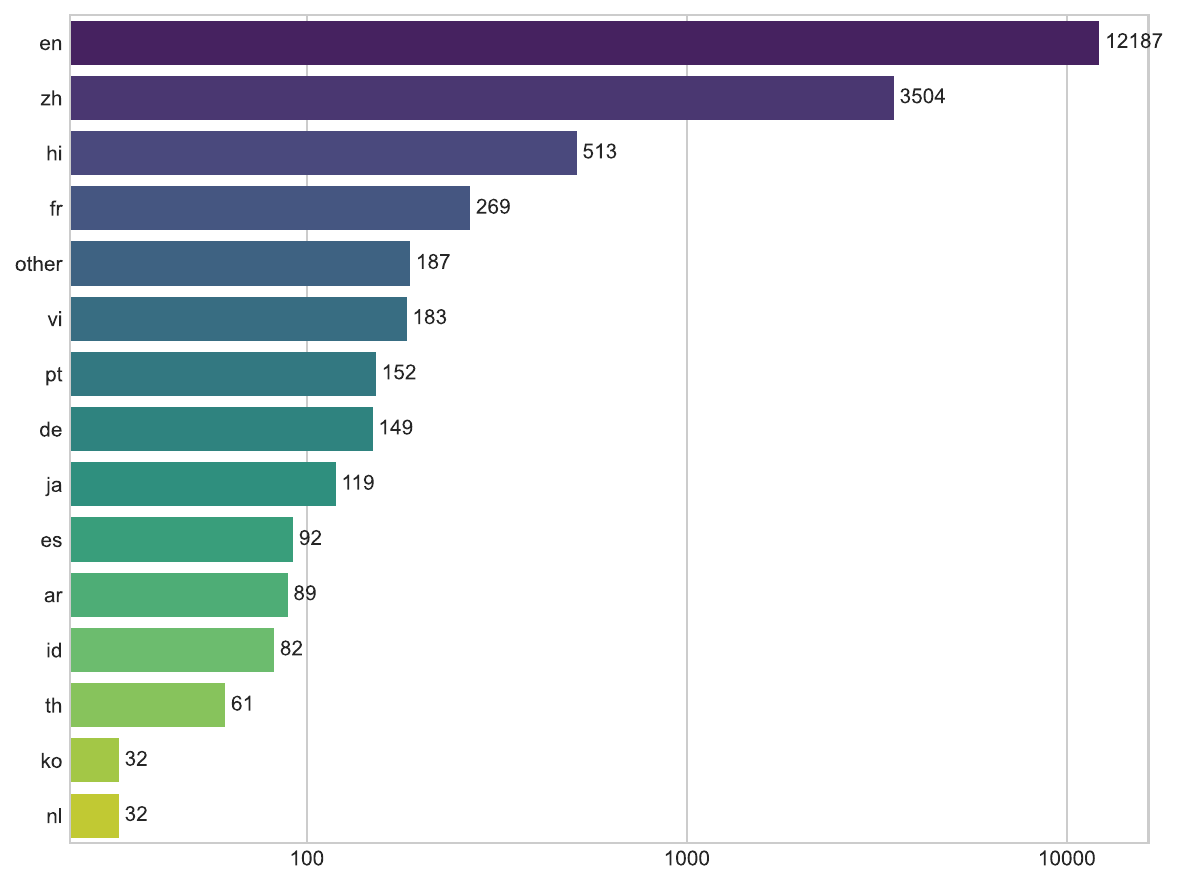}
        \caption{Task language distribution in log scale.}
\label{fig:language}
\end{figure}


APIs are organized into 20 categories (Figure \ref{fig:tool_nums}). The distribution is highly skewed: Finance (14.25\%), Data (12.9\%), Communication (9.75\%), and Entertainment (8.18\%) dominate the dataset. Conversely, categories like Travel (0.22\%), Math (0.84\%), and Sports (0.84\%) are sparsely represented. This disparity highlights a significant concentration in commercial and data-centric domains.

From a geometric perspective, our dataset can be conceptualized in terms of global versus region-specific coverage. Global APIs provide global services—such as machine translation or international weather forecasting—that support multiple languages and are not restricted by geographic boundaries. They make up the majority of our dataset, with 2,291 samples (64.2\%), primarily from providers based in the United States. In contrast, region-specific APIs provide localized services, such as regional weather and news, with major contributions from China and the United States, which together account for 61.79\% of this category. The remaining 38 countries contribute fewer APIs individually due to smaller local markets and less publicly available infrastructure, but their inclusion enhances regional diversity and captures a broader range of localized functionalities worldwide. A detailed distribution is provided in Appendix~\ref{sec:API_Country_Distribution}.


Our dataset consists of 17,540 tasks, including 14,295 single-tool calling tasks and 3,245 multiple-tool calling tasks. The language distribution of all tasks is shown in Figure~\ref{fig:language}. English is the most prevalent language, accounting for 12,187 tasks (69.48\%). This dominance is primarily due to the large proportion of global APIs originating from the United States and the widespread use of English as a lingua franca in API documentation. In addition to English, the dataset contains a rich diversity of 28 other languages. A complete breakdown of all 29 languages and their respective counts is provided in Appendix~\ref{sec:language_dist}.

\section{Experiments and Results}
\label{sec:Experiments_and_Results}
\subsection{Implementation Details}

Our experiments involved both open-source and closed-source large language models (LLMs). The open-source models, which are publicly available for research and development, include general-purpose models such as Qwen2.5~\citep{yang2024qwen2} and DeepSeek-V3~\citep{liu2024deepseek}, as well as models specifically designed for tool calling, such as Hammer2~\citep{lin2024hammer} and Watt-tool-8B. In contrast, the closed-source group comprises state-of-the-art proprietary models, such as GPT-4o, Claude-3.5-Sonnet, and o3-mini. For evaluation on our dataset, open-source models were tested using their default configurations. For fine-tuning, we adopted the LoRA framework~\citep{hu2021lora}, training each model for 3~epochs with a batch size of 1 per device and 8 gradient accumulation steps. The learning rate was set to 1.0e-4, and we employed a cosine learning rate scheduler with a warmup ratio of 0.1. This setup ensures stable convergence while adapting the models to the tool calling tasks in our dataset.

\subsection{Evaluation Metrics}
To comprehensively evaluate model performance, we adopt four evaluation metrics. The first three are based on the Seal-Tools framework~\citep{wu2024seal}: \textbf{(1) Tool Selection (P/R/F1):} Measures the model’s ability to accurately identify the appropriate tool(s) from a set of candidates. Performance is evaluated using precision, recall, and F1-score, reflecting tool localization accuracy; \textbf{(2) Tool Invocation (P/R/F1):} Assesses the model's ability to generate correct and complete tool invocation parameters. We compute precision, recall, and F1 based on triple-level matching of the tool name, parameter key, and parameter value; \textbf{(3) Format Matching Accuracy (FM):} Evaluates whether the model's output conforms to the expected JSON schema. This is a critical requirement for ensuring compatibility with downstream execution environments. 
While these metrics capture key aspects of tool calling, they overlook an essential requirement in multilingual, real-world applications: maintaining linguistic consistency throughout the interaction. To address this gap, we introduce a new metric:
\textbf{(4) Language Matching Accuracy (LM):} Quantifies the proportion of cases in which the model's internal reasoning (i.e., the \textit{thought} field) is expressed in the same language as the user’s input query. We use the \texttt{langid} library for language identification. Detailed formulations and implementation details for all four metrics are provided in Appendix~\ref{sec:Evaluation_Metrics}.

\begin{table}[ht]
\centering
\tiny
\setlength{\tabcolsep}{2pt}
\begin{threeparttable}
\setlength{\tabcolsep}{3pt}  
\begin{tabular}{@{}lcc*{3}{c}*{3}{c}@{}}  
\toprule
\textbf{Model Name} & \textbf{LM} & \textbf{FM} & \multicolumn{3}{c}{\textbf{Tool Selection}} & \multicolumn{3}{c}{\textbf{Tool Invocation}} \\
\cmidrule(lr){4-6} \cmidrule(lr){7-9}
 & & & \textbf{P} & \textbf{R} & \textbf{F1} & \textbf{P} & \textbf{R} & \textbf{F1} \\
\midrule
Qwen2.5-7B-Instruct & 90.51 & 96.65 & 54.08 & 53.06 & 53.18 & 42.76 & 43.37 & 42.71 \\
Qwen2.5-Coder-7B & 94.93 & 98.38 & 69.76 & 66.01 & 67.23 & 54.17 & 54.11 & 53.75 \\
Qwen2.5-3B-Instruct & 87.40 & 93.00 & 49.34 & 45.84 & 47.52 & 40.90 & 41.77 & 41.33 \\
Qwen2.5-Coder-3B & 84.26 & 89.25 & 48.92 & 49.01 & 48.76 & 38.49 & 38.83 & 38.43 \\
watt-tool-8B & 74.48 & 5.53 & \textbf{88.90} & \textbf{88.03} & \textbf{88.30} & \textbf{76.33} & \textbf{73.46} & \textbf{74.31} \\
ToolACE-8B & 81.31 & 4.56 & 70.30 & 69.82 & 69.93 & 59.39 & 56.22 & 57.17 \\
Hammer2.1-7b & 86.82 & 20.71 & 64.64 & 64.68 & 64.44 & 33.14 & 32.68 & 32.75 \\
Hammer2.0-7b & 78.21 & 95.42 & 61.22 & 57.48 & 58.68 & 45.00 & 45.25 & 44.85 \\
Functionary-v3.1 & 76.75 & 54.15 & 40.63 & 37.15 & 38.30 & 35.25 & 35.64 & 35.02 \\
Yi-1.5-9B-Chat-16K & 82.37 & 91.9 & 45.28 & 45.71 & 45.32 & 35.67 & 35.66 & 35.33 \\
glm-4-9b-chat & 76.00 & 97.55 & 43.45 & 41.44 & 42.09 & 32.77 & 32.85 & 32.57 \\
Phi-4 & \textbf{96.73} & 96.29 & 80.90 & 82.68 & 81.49 & 70.15 & 70.25 & 69.84 \\
Qwen2.5-Coder-32B & 91.05 & 99.14 & 84.82 & 81.44 & 82.54 & 71.13 & 71.04 & 70.69 \\
Qwen2.5-72B-Instruct & 89.47 & 98.16 & 52.78 & 51.44 & 51.83 & 43.11 & 43.35 & 42.89 \\
Deepseek-V3 & 86.09 & 99.89 & 83.10 & 83.73 & 83.28 & 75.94 & 75.77 & 75.49 \\
Deepseek-R1 & 77.05 & \textbf{100} & 86.89 & 85.25 & 85.79 & 73.47 & 73.15 & 72.79 \\
\midrule
o1-mini & 95.89 & 93.68 & 64.41 & 66.61 & 64.72 & 60.58 & 62.53 & 61.06 \\
o3-mini & 86.19 & 71.37 & 61.06 & 61.13 & 60.93 & 54.01 & 53.56 & 53.54 \\
GPT4o-mini & 96.24 & \textbf{99.83} & 76.47 & 75.21 & 75.55 & 71.69 & 70.38 & 70.71 \\
GPT4o & \textbf{97.95} & 99.83 & \textbf{88.95} & \textbf{89.48} & \textbf{89.01} & \textbf{82.18} & \textbf{81.57} & \textbf{81.57} \\
GLM-Zero & 88.37 & 98.45 & 51.24 & 50.31 & 50.51 & 42.64 & 43.64 & 42.78 \\
gemini-2.0-flash & 95.04 & 99.77 & 77.25 & 77.76 & 77.32 & 69.08 & 68.14 & 68.18 \\
gemini-2.0-pro & 96.17 & 94.13 & 84.57 & 83.50 & 83.86 & 73.22 & 71.65 & 71.95 \\
Claude-3.5-sonnet & 94.75 & 97.06 & 82.08 & 81.00 & 81.19 & 72.05 & 72.29 & 71.77 \\
\bottomrule
\end{tabular}
\caption{Zero-shot evaluation results on ITC testing data (\%). The best results are highlighted in bold. \vspace{-0.5em}}  
\label{tab:Main results}
\end{threeparttable}
\end{table}

\subsection{Zero-Shot Evaluation of Tool Calling Capabilities}

We evaluate the zero-shot performance of large language models (LLMs) on the ITC test set to assess their intrinsic tool calling capabilities without task-specific fine-tuning.

\noindent\textbf{Overall performance:}
Table~\ref{tab:Main results} presents results across four key metrics: Language Matching (LM), Format Matching (FM), Tool Selection, and Tool Invocation. Overall, closed-source models consistently outperform open-source models. GPT-4o achieves the highest LM (97.95\%) and FM (99.83\%) scores, followed closely by Claude-3.5-Sonnet and Gemini-2.0-Pro. However, the reasoning-focused ``mini'' series (o1-mini and o3-mini) exhibit suboptimal performance despite their strong logic capabilities; their tendency to over-reason on straightforward tool-calling tasks often leads to excessive chain-of-thought generation, which disrupts strict JSON schema adherence (e.g., o3-mini's low FM of 71.37\%) and complicates precise parameter synthesis. Among open-source models, Deepseek-V3 and Qwen2.5-Coder-32B perform well, achieving FM above 99\% and LM above 86\%. In contrast, models such as Watt-tool-8B achieve strong task-level performance but suffer from low LM (74.48\%) and FM (5.53\%), indicating weaknesses in multilingual handling and structural adherence. Lower-performing models like Functionary-v3.1 and Hammer2.1-7B struggle across all dimensions, producing outputs that are often malformed or inconsistent with user language.

\noindent\textbf{Linguistic and structural accuracy:} For LM, most closed-source models exceed 95\%, with GPT-4o at 97.95\%, while open models like Qwen2.5-7B-Instruct (90.51\%) and Phi-4 (96.73\%) also perform well. For FM, Deepseek-R1 reaches 100\%, most closed-source models exceed 95\%, and over two-thirds of open models meet the requirements. Models like Watt-tool-8B and ToolACE-8B have low FM because they generate only tools and parameters without the multi-step reasoning traces required by ITC, causing misformatted or incomplete JSON outputs.

\noindent\textbf{Functional competence in tool calling:}
Closed-source models demonstrate strong overall capabilities in both tool selection and invocation. GPT-4o achieves the highest performance across both tasks, with F1 scores of 89.01\% for Tool Selection and 81.57\% for Tool Invocation. Proprietary models like Gemini-2.0-Pro and Claude-3.5-Sonnet show robust competence with F1 scores exceeding 80\% in both selection and invocation. Among open-source models, Watt-tool-8B leads in tool selection (88.30\% F1) , while DeepSeek-V3 excels in parameter generation (75.49\% F1). Conversely, Hammer2.1-7B and Functionary-v3.1 struggle with invocation (F1 < 36\%), revealing weaknesses in generating executable calls. The performance gap between selection and invocation—most prominent in Hammer2.1-7B—highlights critical challenges in schema adherence and multi-step planning. Such inconsistencies in decision and execution accuracy remain a key barrier to reliable real-world deployment.

\begin{table}[ht]
\centering
\tiny
\setlength{\tabcolsep}{2pt}
\begin{threeparttable}
\setlength{\tabcolsep}{6pt}
\begin{tabular}{@{}l*{3}{c}*{3}{c}@{}}
\toprule
\textbf{Model Name} & \multicolumn{3}{c}{\textbf{Tool Selection}} & \multicolumn{3}{c}{\textbf{Tool Invocation}} \\
\cmidrule(lr){2-4} \cmidrule(lr){5-7}
& \textbf{Hall.} & \textbf{Mis.} & \textbf{Ex.} & \textbf{Incor.} & \textbf{Miss.} & \textbf{Ext.} \\
\midrule
Qwen2.5-7B-Instruct & 21.57 & 73.23 & 5.20 & 51.53 & 19.73 & 28.74 \\
Qwen2.5-Coder-7B & 4.25 & 86.65 & 9.10 & 51.01 & 20.59 & 28.39 \\
Qwen2.5-3B-Instruct & 8.74 & 75.79 & 15.47 & 42.14 & \textbf{16.42} & 41.45 \\
Qwen2.5-Coder-3B & 38.48 & \textbf{51.51} & 10.00 & 37.36 & 23.02 & 39.62 \\
watt-tool-8B & 25.51 & 67.74 & 6.74 & 45.54 & 40.61 & 13.85 \\
ToolACE-8B & 4.12 & 88.75 & 7.13 & 42.03 & 48.63 & \textbf{9.34} \\
Hammer2.1-7b & 0.70 & 91.56 & 7.74 & \textbf{17.18} & 64.26 & 18.56 \\
Hammer2.0-7b & 2.35 & 89.16 & 8.49 & 57.80 & 23.92 & 18.28 \\
Functionary-v3.1 & 20.92 & 76.97 & \textbf{2.11} & 37.70 & 28.80 & 33.51 \\
Yi-1.5-9B-Chat-16K & 37.35 & 55.78 & 6.86 & 38.90 & 18.66 & 42.44 \\
glm-4-9b-chat & 0.98 & 93.90 & 5.12 & 27.88 & 54.78 & 17.34 \\
Phi-4 & 11.29 & 69.18 & 19.53 & 44.75 & 29.11 & 26.14 \\
Qwen2.5-Coder-32B & 8.09 & 66.31 & 25.61 & 46.43 & 27.80 & 25.78 \\
Qwen2.5-72B-Instruct & 43.64 & 51.95 & 4.41 & 46.65 & 22.79 & 30.56 \\
Deepseek-V3 & \textbf{0.43} & 80.21 & 19.36 & 57.65 & 24.38 & 17.97 \\
Deepseek-R1 & 8.33 & 83.33 & 8.33 & 41.67 & 29.17 & 29.17 \\
\midrule
o1-mini & 35.5 & 61.68 & \textbf{2.82} & 54.5 & 28.5 & 17.0 \\
o3-mini & 34.8 & 62.33 & 2.87 & 53.89 & 29.64 & 16.47 \\
GPT4o-mini & 19.71 & 76.26 & 4.03 & 48.76 & 39.94 & \textbf{11.29} \\
GPT4o & 47.16 & \textbf{49.72} & 3.12 & 53.67 & 21.22 & 25.10 \\
GLM-Zero & 37.63 & 58.09 & 4.28 & 48.89 & \textbf{13.27} & 37.83 \\
gemini-2.0-flash & 30.93 & 62.13 & 6.95 & 50.42 & 29.50 & 20.08 \\
gemini-2.0-pro & \textbf{0} & 85.64 & 14.36 & \textbf{47.31} & 35.13 & 17.56 \\
Claude-3.5-sonnet & 22.11 & 68.81 & 9.08 & 54.67 & 21.21 & 24.12 \\
\bottomrule
\end{tabular}
\caption{Error analysis results across different LLMs (\%). Hall.: hallucinating non-existing tools, Mis.: missing required tools, Ex.: calling extra tools, Incor.: generating incorrect parameters, Miss.: missing parameters, Ext.: generating extra parameters. The best results are highlighted in bold.\vspace{-0.5em}}
\label{tab:error_analysis}
\end{threeparttable}
\end{table}


\noindent\textbf{Error Analysis:} Table~\ref{tab:error_analysis} categorizes errors into Selection (hallucination, omission, extra tools) and Invocation (incorrect, missing, or extra parameters). In selection, missing tools is the most frequent error. Gemini-2.0-Pro is highly conservative (0\% hallucination, 85.64\% omission), while GPT-4o is more balanced (47.16\% hallucination, 49.72\% omission). Conversely, open-source models like Qwen2.5-Coder-3B are aggressive but imprecise, with higher hallucination (38.48\%). In invocation, Hammer2.1-7B exemplifies a common failure mode: low incorrect rates (17.18\%) but high missing arguments (64.26\%), often violating API schemas. While Gemini-2.0-Pro and ToolACE-8B show robust, balanced distributions, models like Watt-tool-8B and Functionary-v3.1 tend to over- or under-specify, reflecting weaknesses in schema adherence and planning. These findings underscore that tool omission and parameter errors remain primary obstacles for reliable tool calling.


\subsection{Fine-tuned Evaluation of Tool Calling Capabilities}

In this experiment, We fine-tuned four Qwen2.5 and two DeepSeek models on the ITC training set to assess improvements in tool calling capabilities for open-source LLMs.

\begin{table*}[ht]
\centering
\small
\setlength{\tabcolsep}{1pt}
\begin{threeparttable}
\begin{tabular}{@{}llcc*{3}{c}*{3}{c}@{}}
\toprule
& \textbf{Model} & \textbf{LM} & \textbf{FM} 
& \multicolumn{3}{c}{\textbf{Tool Selection}} 
& \multicolumn{3}{c}{\textbf{Tool Invocation}} \\
\cmidrule(lr){5-7} \cmidrule(lr){8-10}
 & & & & \textbf{P} & \textbf{R} & \textbf{F1} & \textbf{P} & \textbf{R} & \textbf{F1} \\
\midrule
& Qwen2.5-7B-Instruct & 96.9(+6.4) & 99.8(+3.1) & \textbf{97.7}(+43.6) & \textbf{98.1}(+45.0) & \textbf{97.8}(+44.6) & \textbf{90.6}(+47.9) & \textbf{90.6}(+47.2) & \textbf{90.3}(+47.6) \\
& Qwen2.5-Coder-7B & \textbf{97.4}(+2.5) & 99.6(+1.3) & 97.7(+27.9) & 98.0(+32.0) & 97.7(+30.5) & 90.6(+36.4) & 90.4(+36.3) & 90.2(+36.5) \\
& Qwen2.5-3B-Instruct & 97.3(+9.9) & 99.5(+6.5) & 97.4(+48.0) & 97.9(+52.1) & 97.5(\textbf{+50.0}) & 89.8(+48.9) & 89.5(+47.7) & 89.4(+48.0) \\
& Qwen2.5-Coder-3B & 97.3(\textbf{+13.0}) & \textbf{99.8}(+10.5) & 97.6(+48.7) & 97.9(+48.9) & 97.6(+48.9) & 90.3(\textbf{+51.8}) & 90.3(\textbf{+51.4}) & 90.0(\textbf{+51.5}) \\
& DeepSeek-Coder-7B-v1.5 & 77.4(+3.6) & 78.7(+32.5) & 76.5(+51.2) & 76.9(+51.0) & 76.5(+51.2) & 68.4(+48.6) & 68.2(+48.2) & 68.0(+48.4) \\
& DeepSeek-Coder-1.3B & 77.9(+7.9) & 79.3(\textbf{+59.9}) & 56.4(\textbf{+53.1}) & 56.9(\textbf{+53.6}) & 56.4(+53.1) & 46.4(+44.2) & 46.1(+43.9) & 45.9(+43.7) \\
\bottomrule
\end{tabular}
\end{threeparttable}
\caption{Evaluation of fine-tuned models on the ITC test set (\%), with improvements over the original models in brackets.}
\label{tab:results_ft_all}
\end{table*}

\begin{table*}[ht]
\centering
\scriptsize
\setlength{\tabcolsep}{1pt}
\begin{threeparttable}
\begin{tabular}{lcccccccc}
\toprule
\textbf{Model} & \multicolumn{2}{c}{\textbf{Nexus Raven}} & \multicolumn{2}{c}{\textbf{Seal-Tools}} & \multicolumn{2}{c}{\textbf{Tool-Alpaca}} \\
\cmidrule(lr){2-3} \cmidrule(lr){4-5} \cmidrule(lr){6-7}
& \textbf{Tool Selection (P)} & \textbf{Tool Invocation (P)} & \textbf{Tool Selection (P)} & \textbf{Tool Invocation (P)} & \textbf{Tool Selection (P)} & \textbf{Tool Invocation (P)} \\
\midrule
Qwen2.5-7B-Instruct      & 90.57(\textbf{+25.75}) & 59.97(+10.23) & 89.91(\textbf{+24.00}) & 76.16(+17.76) & \textbf{77.05}(\textbf{+18.10}) & 49.96(\textbf{+9.85})  \\
Qwen2.5-Coder-7B-Instruct & \textbf{90.99}(+20.44) & \textbf{68.04}(\textbf{+17.76}) & 89.57(+22.22) & \textbf{78.04}(+18.92) & 77.34(+14.91) & \textbf{50.87}(+8.23)  \\
Qwen2.5-3B-Instruct      & 81.03(+6.50) & 57.14(+2.78) & \textbf{90.32}(+23.26) & 76.76(\textbf{+20.02}) & 75.00(+8.92) & 47.54(+8.57)  \\
Qwen2.5-Coder-3B-Instruct & 84.17(+2.09) & 64.22(+4.90) & 89.34(+8.50) & 76.18(+7.97) & 73.10(+4.53) & 48.69(+7.06)  \\
\bottomrule
\end{tabular}

\caption{Fine-tuned evaluation results on three benchmark testing datasets (\%), with values in brackets showing the improvement from the original models. The best results and greatest improvements are highlighted in bold.}
\label{tab:academic_benchmarks}
\end{threeparttable}
\end{table*}

\begin{table*}[ht!]
\centering
\scriptsize
\begin{threeparttable}
\setlength{\tabcolsep}{2pt}
\begin{tabular}{@{}l@{}cccc*{3}{c}*{3}{c}@{}}
\toprule
\textbf{Type} & \textbf{Model Name} & \textbf{LM} & \textbf{FM} & \multicolumn{3}{c}{\textbf{Tool Selection}} & \multicolumn{3}{c}{\textbf{Tool Invocation}} \\
\cmidrule(lr){5-7} \cmidrule(lr){8-10}
& & & & \textbf{P} & \textbf{R} & \textbf{F1} & \textbf{P} & \textbf{R} & \textbf{F1} \\
\midrule
\multirow{4}{*}{ALL} 
& Qwen2.5-7B-Instruct 
    & 96.30(+5.56)
    & 99.27(+4.18) 
    & 91.57(+36.67) 
    & \textbf{98.57}(\textbf{+42.02}) 
    & \textbf{94.94}(+39.69)
    & 87.78(+45.93) 
    & 86.36(+44.08) 
    & 87.06(+45.45) \\
& Qwen2.5-Coder-7B 
    & \textbf{96.47}(+7.04)
    & \textbf{98.91}(+1.46)
    & \textbf{93.29}(+21.77)
    & 93.55(+19.61) 
    & 93.42(+20.67) 
    & \textbf{88.37}(+33.17) 
    & \textbf{89.19}(+34.80) 
    & \textbf{88.77}(+34.56) \\
& Qwen2.5-3B-Instruct 
    & 91.62(\textbf{+11.24}) 
    & 95.91(\textbf{+13.55})
    & 87.03(+39.58) 
    & 89.76(+40.25) 
    & 88.37(\textbf{+39.91}) 
    & 76.17(+44.06) 
    & 74.23(+41.27) 
    & 75.19(+42.66) \\
& Qwen2.5-Coder-3B 
    & 94.21(+9.34) 
    & \textbf{98.91}(+4.36)
    & 87.44(\textbf{+40.15}) 
    & 86.91(+39.30) 
    & 87.17(+39.86) 
    & 80.20(\textbf{+46.89}) 
    & 80.42(\textbf{+45.66}) 
    & 80.31(\textbf{+46.67}) \\
\midrule
\multirow{4}{*}{EN} & Qwen2.5-7B-Instruct & 91.33(+0.59) &      97.09(+2.00) & 79.55(+24.65) & 79.12(+22.57) & 79.33(+24.08) & 70.82(+28.97) & 71.17(+28.89) & 70.99(+29.38) \\
& Qwen2.5-Coder-7B & \textbf{92.57}(+2.14) & \textbf{98.28}(+1.17)  & \textbf{88.24}(+15.72) & \textbf{88.32}(+14.38) & \textbf{88.28}(+15.53) & \textbf{79.47}(+24.27) & \textbf{79.63}(+25.24) & \textbf{79.55}(+25.34) \\
& Qwen2.5-3B-Instruct & 83.64(\textbf{+3.26}) & 86.51(\textbf{+4.15})  & 77.96(+30.51) & 77.70(+28.19) & 77.83(+29.37) & 69.7(\textbf{+37.59}) & 69.97(\textbf{+37.01}) & 69.83(\textbf{+37.30}) \\
& Qwen2.5-Coder-3B & 85.67(+0.80) & 96.45(+1.90)  & 78.85(\textbf{+31.56}) & 
79.47(\textbf{+31.86}) & 79.16(\textbf{+31.85}) & 69.85(+36.54) & 69.99(+35.23) & 69.92(+36.28) \\
\bottomrule
\end{tabular}
\centering
\caption{Ablation study on non-English queries in the ITC testing dataset evaluating language impact (\%), with values in brackets indicating improvements over the original models. The label ‘Type = ALL’ denotes training on the full ITC dataset, while ‘Type = EN’ indicates training exclusively on the English subset of the ITC dataset. The best results and largest improvements are highlighted in bold.}
\label{tab:Ablation_Study}
\end{threeparttable}
\end{table*}

\noindent\textbf{ITC test set results:} Table~\ref{tab:results_ft_all} shows substantial gains in both tool selection and tool invocation after fine-tuning across all evaluated models. For Qwen, the fine-tuned 3B variants achieve performance comparable to the larger 7B variants. For example, Qwen2.5-7B-Instruct improved tool selection recall by 45.0\% and tool invocation precision by 47.9\%, while Qwen2.5-Coder-3B recorded the largest boost in tool invocation F1 at 51.5\%. These results demonstrate the effectiveness of our training dataset in enhancing tool calling performance across model scales. For DeepSeek, fine-tuning also brings notable gains, with the 7B model outperforming the 1.3B variant across all metrics, achieving up to 51.0\% improvement in tool selection F1 and 48.0\% in tool invocation F1. However, their limited multilingual support and weaker instruction-following leave them trailing the Qwen models on most metrics.

\noindent\textbf{Out-of-domain generalization:} To evaluate robustness beyond the training distribution, we tested the fine-tuned Qwen2.5 models on several external benchmarks (Table~\ref{tab:academic_benchmarks}). All models exhibit marked improvements, with tool selection precision increasing up to 25.8\% and tool invocation precision improving by up to 18.1\%. This indicates that fine-tuning not only strengthens in-domain capabilities but also enhances generalization to unseen tools and tasks.

\subsection{Ablation Study on Language Impact}

To evaluate the impact of non-English languages on model performance, we conducted an ablation study by fine-tuning Qwen2.5 models either on the full multilingual ITC training set (\textit{Type = ALL}) or exclusively on the English subset (\textit{Type = EN}), followed by evaluation on non-English test data. As shown in Table~\ref{tab:Ablation_Study}, models trained on the full dataset achieve substantially higher gains on non-English tasks. For instance, the Qwen2.5-7B-Instruct model fine-tuned on all languages improved tool selection recall by 42.0\%, which is 19.4\% higher than the gain from English-only training. Similarly, tool invocation F1 for Qwen2.5-Coder-7B increased by 34.6\% with full multilingual training, outperforming the English-only gain by 9.3\%. These results indicate that restricting training to English significantly limits performance on non-English tasks, underscoring the importance of incorporating culturally diverse data to enhance LLM generalization in international tool-calling scenarios.


\section{Conclusion}
\label{sec:Conclusion}


In this paper, we introduce the International Tool Calling (ITC) dataset, a geometric diverse and globally representative resource aimed at advancing large language models' (LLMs) capabilities in multi-tool and international API scenarios. Covering a broad range of API categories, ITC addresses critical limitations in existing benchmarks, such as the predominance of English-only queries, insufficient long-tail API coverage, and the lack of complex multi-tool interactions. Our experiments show that fine-tuning on ITC leads to substantial performance gains, including notable improvements on out-of-domain tasks, demonstrating its effectiveness in enhancing LLMs' ability to interact with international APIs.

\section*{Limitations}
\label{sec:Limitations}
While our work presents significant advancements, several limitations warrant further attention. First, despite emphasizing geographical diversity, certain regions (e.g., Africa and parts of Asia) remain underrepresented, potentially limiting the model’s ability to grasp nuanced cultural or regulatory contexts. Second, the dataset focuses solely on REST APIs, leaving other tool types (e.g., SOAP APIs or database connectors) unaddressed, which may constrain applicability in more heterogeneous tool ecosystems. Third, reliance on free APIs introduces potential instability due to service deprecation or rate limits, making regular dataset updates essential to maintain relevance and reproducibility. Finally, more challenging datasets are needed to further boost the tool calling capabilities of open-source LLMs. Addressing these issues will be critical for future work aimed at building truly robust and universal tool calling systems.

\bibliography{custom}

\appendix

\section{API Processing}
\label{sec:API_Processing}

\subsection{API Format}
\begin{figure}[H]
\centering
\tiny
\begin{tcolorbox}[
    title=API Format,
    colback=white,
    colframe=black,
    width=0.9\columnwidth,  
]
\begin{verbatim}
{
    "tool_name":"tool name",
    "tool_description":"tool description",
    "home_url":"home url",
    "country":"Countries involved in the tool",
    "api_list":[
         {
              "name":"api name",
              "url":"api url",
              "description":"api function description",
              "method":"POST/GET",
              "required_parameters":[
                  {
                      "name":"parameter name",
                      "type":"parameter type",
                      "description":"parameter description",
                      "default":"Default value, empty if none"
                   }
              ],
              "optional_parameters":[
                {
                      "name":"parameter name",
                      "type":"parameter type",
                      "description":"parameter description",
                      "default":"Default value, empty if none"
                }
              ],
              "statuscode":"statuscode"
         }
    ]
}
\end{verbatim}
\end{tcolorbox}
\caption{API Format.}
\label{fig:API_Format}
\end{figure}

\subsection{API Error Response}
\begin{figure}[H]
\centering
\includegraphics[width=0.5\columnwidth,keepaspectratio]
  {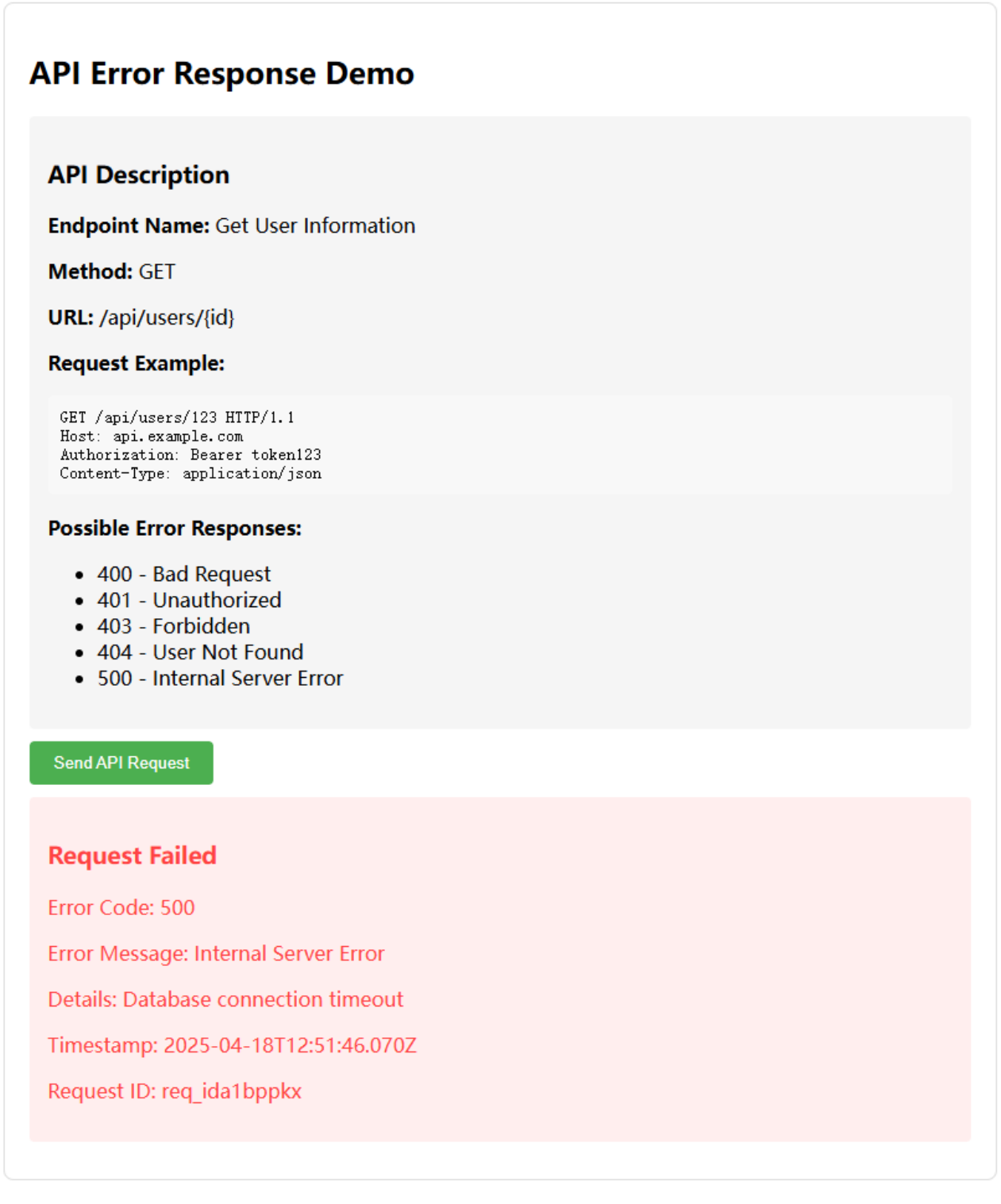}
  \caption{API Error Response Demo.}
  \label{fig:API_Error_Response}
\end{figure}

\subsection{API Empty Response}
\begin{figure}[H]
  \centering
  \includegraphics[width=0.5\columnwidth,keepaspectratio]{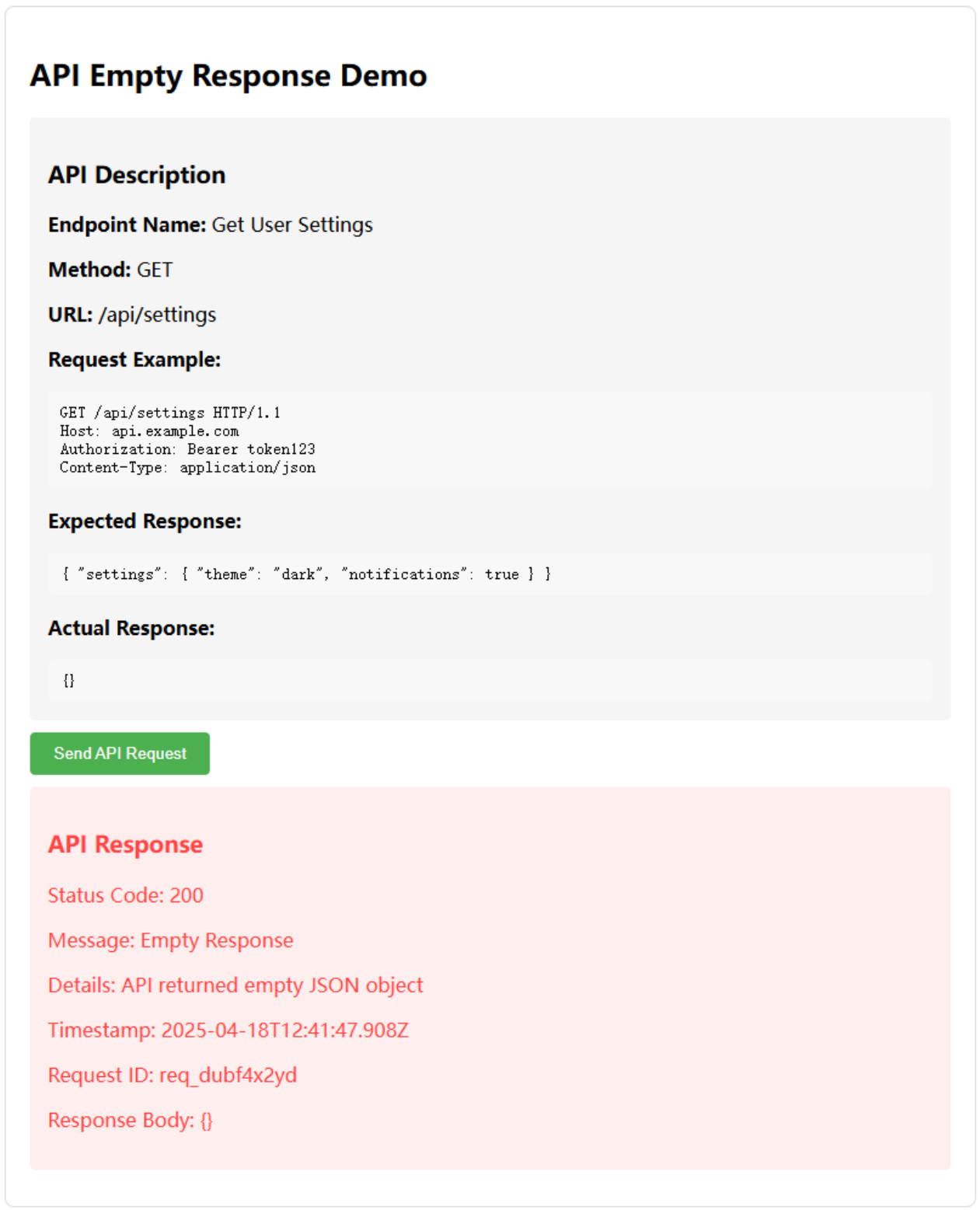}
  \caption{API Empty Response Demo.}
  \label{fig:API_Empty_Response}
\end{figure}





\subsection{API Country Distribution}
\begin{figure}[H]
\centering
\includegraphics[width=\linewidth]{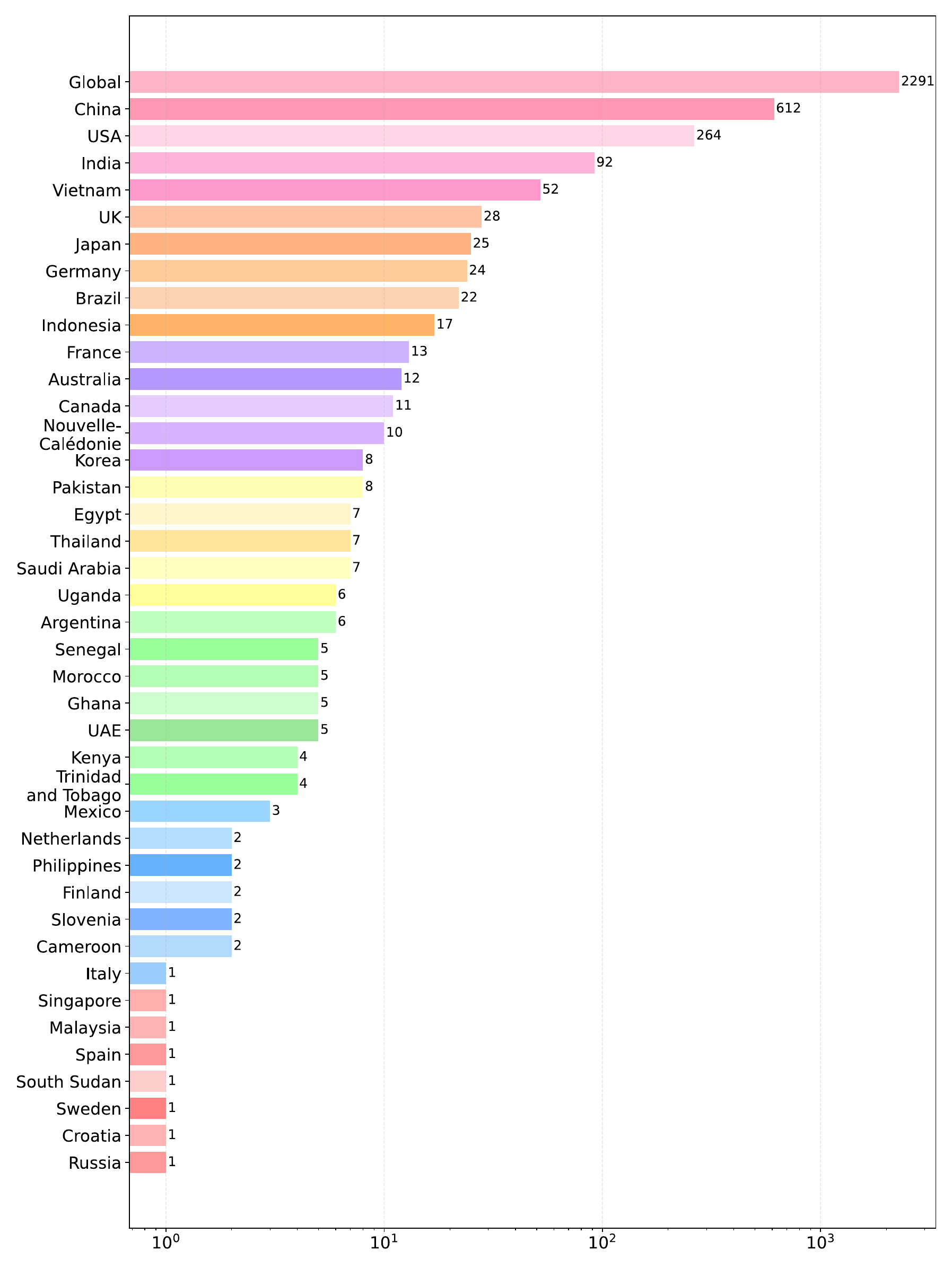}
\caption{Distribution of APIs across countries/regions (log scale).}
\label{fig:country}
\end{figure}

Figure~\ref{fig:country} provides a comprehensive overview of the geographical distribution of APIs in our dataset, including both global and region-specific APIs across more than 30 countries and regions.

\label{sec:API_Country_Distribution}

\section{Detailed Formulate for Evaluation Metrics}
\label{sec:Evaluation_Metrics}
To control page layout, we use \textbf{FM} to represent \textit{Format Matching Accuracy}, \textbf{LM} for \textit{Language Matching Accuracy}, \textbf{Tool} for \textit{Tool Selection}, and \textbf{TI} for \textit{Tool Invocation}.

\begin{equation}
\mathit{LM} = \frac{\mathit{amount}_{\mathit{correct\ language}}}{\mathit{amount}_{\mathit{all}}}
\end{equation}
\begin{equation}
\mathit{FM} = \frac{\mathit{amount}_{\mathit{correct\ format}}}{\mathit{amount}_{\mathit{all}}}
\end{equation}
\begin{equation}
\mathit{Tool\ P} = \frac{\mathit{amount}_{\mathit{correct\ tools}}}{\mathit{amount}_{\mathit{predict\ tools}}}
\end{equation}
\begin{equation}
\mathit{Tool\ R} = \frac{\mathit{amount}_{\mathit{correct\ tools}}}{\mathit{amount}_{\mathit{gold\ tools}}}
\end{equation}
\begin{equation}
\mathit{Tool\ F1} = \frac{2 \cdot \mathit{Tool\ P} \cdot \mathit{Tool\ R}}{\mathit{Tool\ P} + \mathit{Tool\ R}}
\end{equation}
\begin{equation}
\mathit{TI\ P} = \frac{\mathit{amount}_{\mathit{correct\ parameters}}}{\mathit{amount}_{\mathit{predict\ parameters}}}
\end{equation}
\begin{equation}
\mathit{TI\ R} = \frac{\mathit{amount}_{\mathit{correct\ parameters}}}{\mathit{amount}_{\mathit{gold\ parameters}}}
\end{equation}
\begin{equation}
\mathit{TI\ F1} = \frac{2 \cdot \mathit{TI\ P} \cdot \mathit{TI\ R}}{\mathit{TI\ P} + \mathit{TI\ R}}
\end{equation}

\section{Single Tool Calling Tasks Query Generation}
\label{sec:Single_Tool_Calling_Tasks_Query_Generation}
For single tool calling tasks, we utilize a prompt-based approach to instruct the LLM to generate a query. The prompt templates used for this process are illustrated in Figures~\ref{fig:Single_Tool}.
\begin{figure}[ht]
\centering
\small
\begin{tcolorbox}[
    title=Single Tool Calling Tasks Query Generation Prompt,
    colback=white,
    colframe=black,
    width=\linewidth
]

Please strictly follow these guidelines:
1. The instructions should be 1 to 2 sentences long. Use a mix of interrogative sentences, first-person statements, imperative sentences, and other structures that convey a request.Aim for diversity in your instructions.

2. Do not mention the API's name in your instructions.

3. Your instructions should only involve the features provided by these APIs.

4. Generate 10 diverse instructions.

5. Use specific nouns and real-world examples from various domains, such as entertainment, sports, or technology.

6. Please provide concrete details.Don't using any form of generic phrases, such as "this xxx", "the xxx","a xxx" or "a specific xxx".

7. Ensure diversity in language by combining questions with imperative statements and other structures that convey a request.

8. The instructions should be in the language of the country attribute in the provided API information.

9. The generated problem must strictly follow the API's parameter information.

10. If country is Global, please generate 10 instructions in English.
\medskip

Here is the API information: \par
\texttt{\{example\_list\}}

\medskip
Please generate the question in the language of the specified country.

your response:
\end{tcolorbox}
\caption{Query generation prompt for single tool calling tasks.}
\label{fig:Single_Tool}
\end{figure}


\section{Multiple tools Calling Tasks Query Generation}
\label{sec:Multi-tool_Call_Prompt}
\begin{figure}[ht]
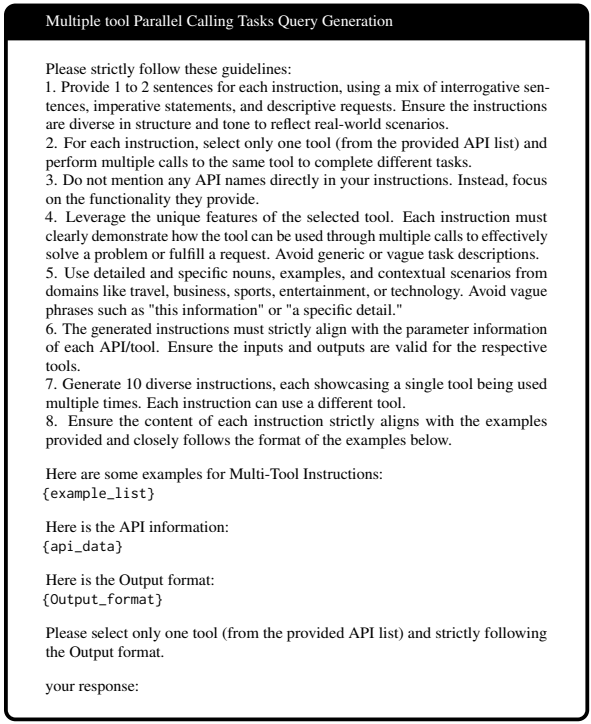

\centering
\tiny
\begin{tcolorbox}[
    title=Multiple tool Parallel Calling Tasks Query Generation,
    colback=white,
    colframe=black,
    width=\columnwidth
]
Please strictly follow these guidelines:

1. Provide 1 to 2 sentences for each instruction, using a mix of interrogative sentences, imperative statements, and descriptive requests. Ensure the instructions are diverse in structure and tone to reflect real-world scenarios.

2. For each instruction, select only one tool (from the provided API list) and perform multiple calls to the same tool to complete different tasks.

3. Do not mention any API names directly in your instructions. Instead, focus on the functionality they provide.

4. Leverage the unique features of the selected tool. Each instruction must clearly demonstrate how the tool can be used through multiple calls to effectively solve a problem or fulfill a request. Avoid generic or vague task descriptions.

5. Use detailed and specific nouns, examples, and contextual scenarios from domains like travel, business, sports, entertainment, or technology. Avoid vague phrases such as "this information" or "a specific detail."

6. The generated instructions must strictly align with the parameter information of each API/tool. Ensure the inputs and outputs are valid for the respective tools.

7. Generate 10 diverse instructions, each showcasing a single tool being used multiple times. Each instruction can use a different tool.

8. Ensure the content of each instruction strictly aligns with the examples provided and closely follows the format of the examples below.
\medskip

Here are some examples for Multi-Tool Instructions: \par
\texttt{\{example\_list\}}

\medskip

Here is the API information: \par
\texttt{\{api\_data\}}

\medskip

Here is the Output format: \par
\texttt{\{Output\_format\}}

\medskip

Please select only one tool (from the provided API list) and strictly following the Output format.

\medskip

your response:
\end{tcolorbox}
\caption{Multiple tool parallel calls.}
\label{fig:Multiple_Tool_Parallel_Calls}
\end{figure}

\begin{figure}[ht]
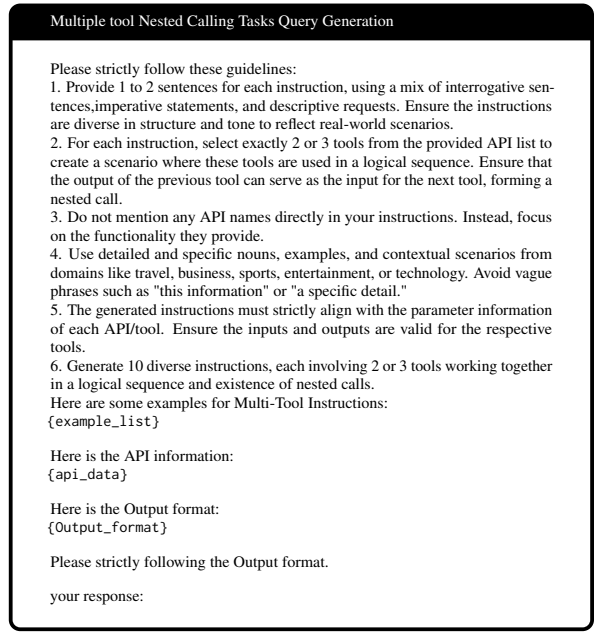

\centering
\tiny
\begin{tcolorbox}[
    title=Multiple tool Nested Calling Tasks Query Generation,
    colback=white,
    colframe=black,
    width=\columnwidth
]
Please strictly follow these guidelines:

1. Provide 1 to 2 sentences for each instruction, using a mix of interrogative sentences,imperative statements, and descriptive requests. Ensure the instructions are diverse in structure and tone to reflect real-world scenarios.

2. For each instruction, select exactly 2 or 3 tools from the provided API list to create a scenario where these tools are used in a logical sequence. Ensure that the output of the previous tool can serve as the input for the next tool, forming a nested call.

3. Do not mention any API names directly in your instructions. Instead, focus on the functionality they provide.

4. Use detailed and specific nouns, examples, and contextual scenarios from domains like travel, business, sports, entertainment, or technology. Avoid vague phrases such as "this information" or "a specific detail."

5. The generated instructions must strictly align with the parameter information of each API/tool. Ensure the inputs and outputs are valid for the respective tools.

6. Generate 10 diverse instructions, each involving 2 or 3 tools working together in a logical sequence and existence of nested calls.

Here are some examples for Multi-Tool Instructions:

\texttt{\{example\_list\}}

\medskip
Here is the API information:\par
\texttt{\{api\_data\}}

\medskip
Here is the Output format:\par
\texttt{\{Output\_format\}}

\medskip
Please strictly following the Output format.\par
\medskip
your response:
\end{tcolorbox}
\caption{Multiple tool nested calls.}
\label{fig:Multiple_Tool_Nested_Calls}
\end{figure}

For multiple tool calling tasks, we have classified them into three categories: Repeated Calls, Parallel Calls, and Nested Calls. Given that the requirements for each type of task differ, we have tailored specific prompts to generate queries for each category. The prompt templates for these tasks are illustrated in Figures~\ref{fig:Multiple_Tool_Repeated_Calls}, \ref{fig:Multiple_Tool_Parallel_Calls}, and \ref{fig:Multiple_Tool_Nested_Calls}.

\begin{figure}[ht]
\centering
\tiny
\begin{tcolorbox}[
    title=Multiple tool Repeated Calling Tasks Query Generation,
    colback=white,
    colframe=black,
    width=\columnwidth    
]
Please strictly follow these guidelines:
1. Provide 1 to 2 sentences for each instruction, using a mix of interrogative sentences, imperative statements, and descriptive requests. Ensure the instructions are diverse instructure and tone to reflect real-world scenarios.

2. For each instruction, select only one tool (from the provided API list) and perform multiple calls to the same tool to complete different tasks.

3. Do not mention any API names directly in your instructions. Instead, focus on the functionality they provide.

4. Leverage the unique features of the selected tool. Each instruction must clearly demonstrate how the tool can be used through multiple calls to effectively solve a problem or fulfill a request. Avoid generic or vague task descriptions.

5. Use detailed and specific nouns, examples, and contextual scenarios from domains like travel, business, sports, entertainment, or technology. Avoid vague phrases such as "this information" or "a specific detail."

6. The generated instructions must strictly align with the parameter information of each API/tool. Ensure the inputs and outputs are valid for the respective tools.

7. Generate 10 diverse instructions, each showcasing a single tool being used multiple times. Each instruction can use a different tool.

8. Ensure the content of each instruction strictly aligns with the examples provided and closely follows the format of the examples below.
\medskip

Here are some examples for Multi-Tool Instructions:\par
\texttt{\{example\_list\}}
\medskip

Here is the API information:\par
\texttt{\{api\_data\}}

\medskip
Here is the Output format:\par
\texttt{\{Output\_format\}}

\medskip
Please select only one tool (from the provided API list) and strictly 
following the Output format.
\medskip

your response:
\end{tcolorbox}
\caption{Multiple tool repeated calls.}
\label{fig:Multiple_Tool_Repeated_Calls}
\end{figure}

\section{Query Scoring}
\label{sec:Example_of_scoring}

\subsection{Scoring dimensions}
\label{sec:Scoring_dimensions}
To comprehensively assess the quality of instructions (queries or question-and-answer pairs), we adopt the following five evaluation dimensions:
\begin{enumerate}
    \item \textbf{Relevance}: Measures the alignment between the instruction and the task objective. High-scoring instructions accurately reflect the task requirements, while irrelevant or off-topic instructions receive lower scores.
    \item \textbf{Practicality}: Assesses the feasibility and executability of the instruction in real-world scenarios. High scores indicate instructions that can be directly implemented without significant obstacles.
    \item \textbf{Linguistic Applicability}: Evaluates the instruction’s adherence to grammatical norms and consideration of cultural and linguistic context. High-scoring instructions are well-phrased, natural, and unambiguous.
    \item \textbf{Clarity}: Judges whether the instruction is clearly articulated, logically coherent, and easy to understand. High scores indicate concise, explicit, and actionable instructions.
    \item \textbf{Specificity}: Measures the level of detail and focus in the instruction. High-scoring instructions clearly define the scope of operation, reduce ambiguity, and facilitate precise tool invocation.
\end{enumerate}
Each dimension is scored on a scale from 1 to 5, where 1 indicates very low quality and 5 indicates very high quality. The detailed scoring criteria are shown in Table~\ref{tab:scoring_guidelines}

\begin{table*}[ht]
\caption{Scoring guidelines for each evaluation dimension.}
\label{tab:scoring_guidelines}
\centering
\small
\begin{tabularx}{\textwidth}{|l|X|X|X|}
\toprule
\textbf{Dimension} & \textbf{1 (Very Low)} & \textbf{3 (Medium)} & \textbf{5 (Very High)} \\
\midrule
Relevance & Completely irrelevant & Partially relevant & Highly aligned with the task \\
Practicality & Hard to implement & Feasible but with obstacles & Directly applicable \\
Linguistic Applicability & Incorrect or awkward & Basically fluent & Standard, natural, idiomatic \\
Clarity & Vague or confusing & Understandable but ambiguous & Clear and unambiguous \\
Specificity & Overly broad or vague & Some details present & Detailed, focused, unambiguous \\
\bottomrule
\end{tabularx}
\end{table*}

\subsection{Crowdsourced Annotation Quality Control}
\label{sec:annotation_details}

To validate the automated scoring results and ensure the dataset's reliability, we conducted a large-scale human verification campaign. The specific implementation details and quality metrics are as follows.

\paragraph{Worker Qualification and Recruitment.}
Our recruitment pipeline prioritized domain expertise and linguistic proficiency. The selection process involved three strict stages:
\begin{itemize}
    \item \textbf{Stage 1: Linguistic \& Logic Screening.} Candidates were tested on their native-level proficiency in the target language and their ability to understand complex API logic.
    \item \textbf{Stage 2: Guideline Comprehension Exam.} Applicants took an exam based on the scoring dimensions defined in Section~\ref{sec:Scoring_dimensions}.
    \item \textbf{Stage 3: Pilot Qualification.} Candidates annotated a batch of 20 pre-labeled queries. Only those achieving $\geq$ 90\% accuracy against the expert ground truth were qualified.
\end{itemize}
From an initial pool of applicants, we recruited 100 qualified annotators to perform the final verification.

\paragraph{Real-time Quality Monitoring.}
To maintain high standards during the large-scale annotation, we employed a "Gold Standard" injection method. We embedded 1,500 expert-verified queries (sentinels) randomly into the task stream, constituting approximately 10\% of the total workload.
\begin{itemize}
    \item Annotators were unaware of which queries were sentinels.
    \item Workers whose accuracy on these sentinel items dropped below 85\% were automatically flagged, their recent work was discarded, and they were removed from the project.
\end{itemize}

\paragraph{Inter-Annotator Agreement (IAA).}
Each query was reviewed by at least two independent annotators. We calculated Fleiss' $\kappa$ to evaluate the consistency of human judgments. As shown in Table~\ref{tab:full_agreement}, we achieved an overall $\kappa$ of 0.68, indicating substantial agreement.

\begin{table}[ht]
\centering
\small
\setlength{\tabcolsep}{8pt} 
\begin{tabularx}{\columnwidth}{Xcc} 
\toprule
\textbf{Dimension} & \textbf{Fleiss' $\kappa$} & \textbf{Agreement} \\
\midrule
Relevance                & 0.72 & Substantial \\
Practicality             & 0.65 & Substantial \\
Linguistic Applicability & 0.74 & Substantial \\
Clarity                  & 0.69 & Substantial \\
Specificity              & 0.61 & Moderate    \\
\midrule
\textbf{Overall}         & \textbf{0.68} & \textbf{Substantial} \\
\bottomrule
\end{tabularx}
\caption{Detailed IAA statistics.}
\label{tab:full_agreement}
\end{table}

The dimension \textit{Linguistic Applicability} showed the highest agreement ($\kappa=0.74$), confirming the effectiveness of our native-speaker requirement. \textit{Specificity} showed moderate agreement ($\kappa=0.61$), reflecting the inherent subjectivity in judging granular API requirements; in these cases, expert adjudication was used to resolve disagreements.

\subsection{Example of scoring}
\begin{figure}[ht] 
\centering
  \includegraphics[width=0.9\linewidth]{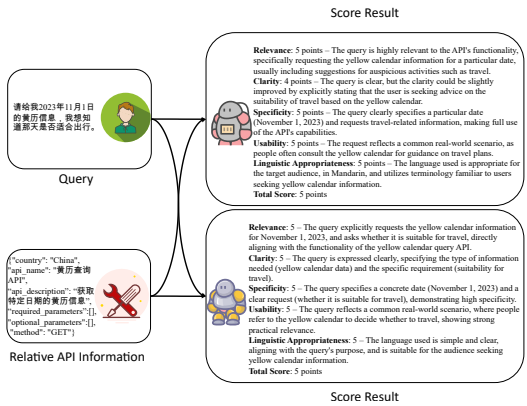} 
  \caption{The query scoring process.}
  \label{fig:score}
\end{figure}

Figure~\ref{fig:score} illustrates an example of query scoring, where, given a query and relevant API information, we used both Anthropic’s Claude-3.5-sonnet model and Google’s Gemini-1.5-pro model to evaluate the query's quality across five dimensions, with scores ranging from 1 to 5 for each dimension. Figure~\ref{fig:Scoring_Criteria} shows the prompt for LLMs to evaluate the query.

\begin{figure}[ht]
\centering
\footnotesize
\begin{tcolorbox}[
    title=Query Scoring Prompt,
    colback=white,
    colframe=black,
    width=\columnwidth
]
Evaluation Criteria:
Use a 1-5 scale to score the following five dimensions:

1. Relevance: How well the query matches the API's functionality.

2. Clarity: Whether the query is specific enough, avoiding ambiguous terms like 'this xxx', 'the xxx', or 'a xxx', and ensuring the use of the API's features.

3. Specificity: Whether the query is specific enough to utilize the API's capabilities

4. Practicality: Whether the query reflects real-world usage scenarios

5. Language Appropriateness: Whether the query's language is suitable for target users

\medskip
Scoring Standard:

1 point: Does not meet the standard

2 points: Partially meets the standard

3 points: Meets the basic standard

4 points: Meets the standard well

5 points: Fully meets the standard

\medskip

Total Score Calculation:

Calculate the average of the five dimension scores, round to the nearest integer, as the final total score (1-5 points).

\medskip
Evaluation Steps:

1. Carefully read the API name and the generated query.

2. Score each dimension and provide a brief explanation.

3. Calculate the total score.

4. Provide an overall evaluation and suggestions for improvement.

5. If the total score is less than 3, mark it as "Needs Improvement".

\medskip
Output Format:

Scores:

1. Relevance: [Score] - [Explanation]

2. Clarity: [Score] - [Explanation]

3. Specificity: [Score] - [Explanation]

4. Practicality: [Score] - [Explanation]

5. Language Appropriateness: [Score] - [Explanation]

Total Score: [1-5 points]

\medskip
Overall Evaluation:

[Brief summary of the query's strengths and weaknesses]

\medskip
Improvement Suggestions:

[Provide specific suggestions for improvement if needed]

\medskip
Conclusion: [If total score >= 4, then "Pass"; if total score < 4, then "Needs Improvement"]
Please evaluate the following data's query{data['query']}

\medskip
Your response:
\end{tcolorbox}
\caption{Query scoring prompt.}
\label{fig:Scoring_Criteria}
\end{figure}

\section{QA Generation}
\label{sec:QA_Generation}
To further evaluate the model’s ability to employ APIs as external tools in multilingual 
settings, we design a dedicated \textit{QA Generation} prompt.Specifically, the response format requires two components: 
\texttt{Thought}, which captures the intermediate reasoning steps, and \texttt{Action}, 
which specifies the chosen API call along with the necessary parameters. Additionally, 
the model is instructed to answer strictly in the language specified by the provided 
\texttt{country} attribute, ensuring robustness in multilingual environments. The complete 
prompt template is presented in Figure~\ref{fig:QA_Generation_Prompt}.

\begin{figure}[ht]
\centering
\footnotesize
\begin{tcolorbox}[
    title=QA Generation Prompt,
    colback=white,
    colframe=black,
    width=\columnwidth
]
You are an expert in using APIs as tools and are highly knowledgeable in multilingual environments. For each task I provide, along with the corresponding API information, you will use the API to complete it.

Your response should follow this format:

\begin{lstlisting}
```json
{
  "Thought": "Your thought process when facing a problem.",
  "Action": "The API you chose and its parameter information. For example [api_name(parameter_1=\"\",parameter_2=\"\"),api_name(parameter_1=\"\",parameter_2=\"\")]"
}
```
\end{lstlisting}
Please strictly respond in the language specified by the \{country\} attribute in the API information.

\medskip
The API information is as follows:\par
\texttt{\{api\_info\}}

\medskip
The task is as follows:\par
\texttt{\{query\}}

\medskip
Your response:
\end{tcolorbox}
\caption{Query scoring prompt.}
\label{fig:QA_Generation_Prompt}
\end{figure}

\section{Checker}
\label{sec:Checker}
To ensure data quality during the multi-turn QA pair generation process, we designed and introduced an LLM-based Checker module for quality filtering. This module is used to determine whether automatically generated QA pairs meet the following criteria:

\begin{itemize}
    \item \textbf{Consistency}: Whether the question and answer are semantically aligned, and whether the answer is genuinely based on the API response.
    \item \textbf{Reasonability}: Whether the answer reasonably reflects the tool's output and avoids fabrication.
    \item \textbf{Linguistic Quality}: Whether the sentence is fluent and grammatically correct.
\end{itemize}

\subsection{Implementation Details of the Checker}
We used Claude-3.5-sonnet and Gemini-1.5-pro as the primary quality assessment model. The prompt is as illustrated in Figures~\ref{fig:Checker_Prompt}. We set the temperature of the Checker to 0 to ensure stability in its judgments.

\begin{figure}[ht]
\centering
\footnotesize
\begin{tcolorbox}[
title=Checker Prompt,
colback=white,
colframe=black,
width=\columnwidth
]

You are an advanced AI assistant acting as a strict Quality Checker for a multi-turn, tool using QA dataset. Your primary function is to evaluate the quality of the final assistant's answer in a given dialog.

\medskip
Your task is to analyze the provided dialog. You must determine if the final "assistant" answer meets our quality standards based on the preceding user question and the tool's response in that turn. Your judgment must be strict. If you find any issue, you must mark it as "Fail".

\medskip
You will be provided with the following information:

1.  **Tool List**: A list of available tools. This provides context on the capabilities the assistant could use.

2.  **Dialog**: The complete conversation flow, formatted as a list of turns. 

3.  **Example List**: A list of pre-judged examples ("Pass" or "Fail"). Use these examples to understand and calibrate your judgment according to our quality standards. Do NOT evaluate the examples themselves.

\medskip
The final assistant's answer must satisfy ALL of the following criteria to be marked as "Pass".

1.  **Consistency**:
    -   **Semantic Alignment**: Does the answer directly and relevantly address the user's latest question?
    -   **Factual Grounding**: Is the answer genuinely and exclusively based on the information from the tool's API response in that turn? It must not contradict the data or omit critical requested information.

2.  **Reasonability**:
    -   **No Fabrication**: The answer must NOT contain any information, details, or suggestions that are not present in the API response. It must not hallucinate facts.
    -   **Accurate Reflection**: Does the answer accurately summarize or present the tool's output without exaggeration or misinterpretation?

3.  **Linguistic Quality**:
    -   **Fluency \& Grammar**: Is the answer fluent, natural-sounding, and grammatically correct? It should be free of awkward phrasing and errors.

\medskip
Your entire response MUST be a single JSON object with two keys, Here is the Output format:

\begin{lstlisting}
```json
{
    "decision": "A string, either \"Pass\" or \"Fail\"",
    "reasoning": "A brief string explaining your decision, referencing the specific criterion that was violated or confirming that all criteria were met."
}
```
\end{lstlisting}

Here are examples to guide your judgment.

\{example\_list\}

\medskip
Here is a list of tools that will be used in the dialog:
\{tool\_list\}

\medskip
Here is the full dialog:
\{dialog\}

\medskip
your response:
\end{tcolorbox}
\caption{Checker Prompt.}
\label{fig:Checker_Prompt}
\end{figure}

\section{Full Language Distribution}
\label{sec:language_dist}
Table~\ref{tab:full_language_distribution} provides the complete list of all 29 languages present in our dataset, along with the exact number of tasks for each language.

\begin{table}[h!]
  \centering
  \caption{Complete distribution of all 29 languages and their task counts.}
  \label{tab:full_language_distribution}
  \begin{tabular}{l r | l r}
    \toprule
    \textbf{Language} & \textbf{Count} & \textbf{Language} & \textbf{Count} \\
    \midrule
    en & 12,187 & sl & 24 \\
    zh & 3,504  & xh & 22 \\
    hi & 513    & ur & 22 \\
    fr & 269    & cs & 21 \\
    vi & 183    & la & 16 \\
    pt & 152    & tl & 11 \\
    de & 149    & ms & 9 \\
    ja & 119    & it & 9 \\
    es & 92     & rw & 8 \\
    ar & 89     & fi & 6 \\
    id & 82     & hr & 3 \\
    th & 61     & gl & 3 \\
    ko & 32     & kn & 2 \\
    nl & 32     & zu & 2 \\
    sw & 29     &    &    \\ 
    \bottomrule
  \end{tabular}
\end{table}

\section{Data Examples}
\label{sec:Example_data}

\begin{figure}[ht]
\centering
\footnotesize
\begin{tcolorbox}[
    title=API Example,
    colback=white,
    colframe=black,
    width=\columnwidth
]
\begin{lstlisting}
{
  "tool_name": "Google Translate",
  "tool_description": "A tool for translating text between different languages using Google's translation services.",
  "home_url": "https://rapidapi.com/nickrapidapi/api/google_translate/",
  "country": "Global",
  "api_list": [
    {
      "name": "translate1",
      "url": "https://google_translate.p.rapidapi.com/",
      "description": "This API endpoint allows users to perform translations of text from one language to another.",
      "method": "GET",
      "category": "Utilities",
      "required_parameters": [
        {
          "name": "text",
          "type": "string",
          "description": "The text content to be translated"
        },
        {
          "name": "target_lang",
          "type": "string",
          "description": "The target language code (e.g., 'en' for English, 'zh' for Chinese)"
        }
      ],
      "optional_parameters": [],
      "statuscode": "200",
      "schema": {}
    }
  ]
}
\end{lstlisting}
\end{tcolorbox}
\caption{Example of Google Translate API.}
\label{fig:example_api}
\end{figure}

Figure~\ref{fig:example_api} illustrates an example of the Google Translate API. Figure~\ref{fig:Single_Tool_Calling_Task_Example} provides an example of a single tool calling task, while Figure~\ref{fig:Multi_Tool_Repeated_Calling_Task_Example} demonstrates a repeated multiple tools calling task. Figure~\ref{fig:Multi_Tool_Parallel_Calling_Task_Example} shows an example of a parallel multiple tools calling task, and Figure~\ref{fig:Multi_Tool_Nested_Calling_Task_Example} presents an example of a nested multiple tools calling task.

\begin{figure}[ht]
\centering
\footnotesize
\begin{tcolorbox}[
title=Single Tool Calling Task Example,
colback=white,
colframe=black,
width=\linewidth
]
\begin{lstlisting}
{
"system": "You are an advanced AI assistant with expertise in:
1. Intelligent tool and API selection
2. Multilingual problem-solving
3. Systematic analytical thinking
Task Guidelines:
        - Carefully analyze the problem presented
        - Select the most appropriate tool(s) or API(s) to solve the task
        - Provide clear, structured reasoning for your tool selection
        - Ensure precise and efficient problem resolution

Important Principles:
        - If no suitable tool exists, clearly explain the limitations
        - Prioritize accuracy and relevance in tool selection
        - Consider potential alternative approaches
        - Maintain a logical and methodical approach

When tools are insufficient or parameters are incomplete:
        - Explicitly state the gaps in available tools or information
        - Provide constructive suggestions for obtaining resources
        - Return an empty list in the \"Action\" section
Note: Please respond in the same national language as the problem
while maintaining clarity, logic, and structure in your answers.
Below is the list of functions you can invoke (in JSON format):
{api_list}

Please strictly follow the format below, without additional text:
        
json{\"Thought\": \"Respond in same language as problem. Provide 
          analysis, reasoning for tool selection, and challenges.\",
          \"Action\": \"[function_name1(param_1='val1', param_2='val2'), 
          function_name2(param_1='val1', param_2='val2')]\"}",
"conversations": [
{
"from": "user",
"value": "Can you help me translate "Hello, how are you?" into French for my 
upcoming trip to Paris?"
},
{
"from": "assistant",
"value": "json{\"Thought\":\"To assist the user with their request for 
translation, I will utilize the Google Translate API, which is specifically 
designed for translating text between languages. The user needs a translation 
for a common greeting in French, which is straightforward.\",
\"Action\":\"[Google Translate(text='Hello, how are you?',"target_lang"="fr")]
\"}"}]
}
\end{lstlisting}
\end{tcolorbox}
\caption{Single tool calling task example.}
\label{fig:Single_Tool_Calling_Task_Example}
\end{figure}

\begin{figure}[ht]
\centering
\footnotesize
\begin{tcolorbox}[
title=Repeated Multiple Tools Calling Task Example,
colback=white,
colframe=black,
width=\linewidth
]
\begin{lstlisting}
{
"system": "You are an advanced AI assistant with expertise in:
1. Intelligent tool and API selection
2. Multilingual problem-solving
3. Systematic analytical thinking
Task Guidelines:
        - Carefully analyze the problem presented
        - Select the most appropriate tool(s) or API(s) to solve the task
        - Provide clear, structured reasoning for your tool selection
        - Ensure precise and efficient problem resolution

Important Principles:
        - If no suitable tool exists, clearly explain the limitations
        - Prioritize accuracy and relevance in tool selection
        - Consider potential alternative approaches
        - Maintain a logical and methodical approach

When tools are insufficient or parameters are incomplete:
        - Explicitly state the gaps in available tools or information
        - Provide constructive suggestions for obtaining resources
        - Return an empty list in the \"Action\" section
Note: Please respond in the same national language as the problem
while maintaining clarity, logic, and structure in your answers.
Below is the list of functions you can invoke (in JSON format):
{api_list}

Please strictly follow the format below, without additional text:
        
json{\"Thought\": \"Respond in same language as problem. Provide 
          analysis, reasoning for tool selection, and challenges.\",
          \"Action\": \"[function_name1(param_1='val1', param_2='val2'), 
          function_name2(param_1='val1', param_2='val2')]\"}",
"conversations": [
{
"from": "user",
"value": "Can you verify the email address 'john.doe@example.com' using the 
Express option? After that, check 'jane.smith@sample.com' with the Premium 
option to see if it's valid."
},
{
"from": "assistant",
"value": "json{\"Thought\":\"To verify the email addresses, I will use the 
Global Email V4 API. First, I will check 'john.doe@example.com' with the Express 
option, and then I will verify 'jane.smith@sample.com' using the Premium option.",
\"Action\":\"[Global Email V4(opt="VerifyMailbox:Express", email=
"john.doe@example.com"),Global Email V4(opt="VerifyMailbox:ExpressPremium", 
email="jane.smith@sample.com")]
\"}"}]
}
\end{lstlisting}
\end{tcolorbox}
\caption{Repeated multiple tools calling task example.}
\label{fig:Multi_Tool_Repeated_Calling_Task_Example}
\end{figure}

\begin{figure}[ht]
\centering
\footnotesize
\begin{tcolorbox}[
title=Parallel Multiple Tools Calling Task Example,
colback=white,
colframe=black,
width=\linewidth
]
\begin{lstlisting}
{
"system": "You are an advanced AI assistant with expertise in:
1. Intelligent tool and API selection
2. Multilingual problem-solving
3. Systematic analytical thinking
Task Guidelines:
        - Carefully analyze the problem presented
        - Select the most appropriate tool(s) or API(s) to solve the task
        - Provide clear, structured reasoning for your tool selection
        - Ensure precise and efficient problem resolution

Important Principles:
        - If no suitable tool exists, clearly explain the limitations
        - Prioritize accuracy and relevance in tool selection
        - Consider potential alternative approaches
        - Maintain a logical and methodical approach

When tools are insufficient or parameters are incomplete:
        - Explicitly state the gaps in available tools or information
        - Provide constructive suggestions for obtaining resources
        - Return an empty list in the \"Action\" section
Note: Please respond in the same national language as the problem
while maintaining clarity, logic, and structure in your answers.
Below is the list of functions you can invoke (in JSON format):
{api_list}

Please strictly follow the format below, without additional text:
        
json{\"Thought\": \"Respond in same language as problem. Provide 
          analysis, reasoning for tool selection, and challenges.\",
          \"Action\": \"[function_name1(param_1='val1', param_2='val2'), 
          function_name2(param_1='val1', param_2='val2')]\"}",
"conversations": [
{
"from": "user",
"value": "Can you check the latest exchange rates for USD and AUD against EUR, 
and also give me a list of all available currencies?"
},
{
"from": "assistant",
"value": "json{\"Thought\":\"To fulfill the request, I need to retrieve the 
latest exchange rates for USD and AUD against EUR, as well as a list of all 
available currencies. This requires using two different API endpoints: one for 
the latest exchange rates and another for the list of currencies.",
\"Action\":\"[latest(from='EUR', to='USD,AUD'), currencies()]
\"}"}]
}
\end{lstlisting}
\end{tcolorbox}
\caption{Parallel multiple tools calling task example.}
\label{fig:Multi_Tool_Parallel_Calling_Task_Example}
\end{figure}

\begin{figure}[ht]
\centering
\footnotesize
\begin{tcolorbox}[
title=Nested Multiple Tools Calling Task Example,
colback=white,
colframe=black,
width=\linewidth
]
\begin{lstlisting}
{
"system": "You are an advanced AI assistant with expertise in:
1. Intelligent tool and API selection
2. Multilingual problem-solving
3. Systematic analytical thinking
Task Guidelines:
        - Carefully analyze the problem presented
        - Select the most appropriate tool(s) or API(s) to solve the task
        - Provide clear, structured reasoning for your tool selection
        - Ensure precise and efficient problem resolution

Important Principles:
        - If no suitable tool exists, clearly explain the limitations
        - Prioritize accuracy and relevance in tool selection
        - Consider potential alternative approaches
        - Maintain a logical and methodical approach

When tools are insufficient or parameters are incomplete:
        - Explicitly state the gaps in available tools or information
        - Provide constructive suggestions for obtaining resources
        - Return an empty list in the \"Action\" section
Note: Please respond in the same national language as the problem
while maintaining clarity, logic, and structure in your answers.
Below is the list of functions you can invoke (in JSON format):
{api_list}

Please strictly follow the format below, without additional text:
        
json{\"Thought\": \"Respond in same language as problem. Provide 
          analysis, reasoning for tool selection, and challenges.\",
          \"Action\": \"[function_name1(param_1='val1', param_2='val2'), 
          function_name2(param_1='val1', param_2='val2')]\"}",
"conversations": [
{
"from": "user",
"value": "Can you generate a secure password with numbers and then hash that 
password using the MD5 algorithm?"
},
{
"from": "assistant",
"value": "json{\"Thought\":\"To generate a secure password with numbers, I will 
use the 'Generate password' API with the appropriate parameters. After 
generating the password, I will hash it using the 'MD5 Text Hash' API.",
\"Action\":\"[Generate password(numbers=true,pwCount=1,length=12),MD5 Text 
Hash(dataString='generated_password')]
\"}"
},
{
"role": "Tool",
"content": {\"code\": 200,\"msg\": \"Password generated successfully.\",
\"data\": {\"password\": \"A1b2C3d4E5f6\"    }}"
},
{
"from": "assistant",
"value": "json{\"Thought\":\"I have generated the password 'A1b2C3d4E5f6' 
successfully. Now, I will hash this password using the MD5 algorithm.",
\"Action\":\"[MD5 Text Hash(dataString='A1b2C3d4E5f6')]
\"}"
}
]
}
\end{lstlisting}
\end{tcolorbox}
\caption{Nested multiple tools calling task example.}
\label{fig:Multi_Tool_Nested_Calling_Task_Example}
\end{figure}

\end{document}